\documentclass[]{aa}
\usepackage{txfonts,epsfig}
\begin{document}
\title{Probing the Local Bubble with Diffuse Interstellar Bands.\\ I. Project
overview and Southern hemisphere survey}
\author{Mandy Bailey\inst{1,2}
   \and Jacco Th.\ van Loon\inst{1}
   \and Amin Farhang\inst{3}
   \and Atefeh Javadi\inst{3}
   \and Habib G.\ Khosroshahi\inst{3}
   \and Peter J.\ Sarre\inst{4}
   \and Keith T.\ Smith\inst{4,5}}
\institute{Lennard-Jones Laboratories, Keele University, ST5 5BG, UK\\
           \email{mandybailey22@gmail.com}
\and       Astrophysics Research Institute, Liverpool John Moores University,
           IC2, Liverpool Science Park, Liverpool L3 5RF, UK
\and       School of Astronomy, Institute for Research in Fundamental Sciences
           (IPM), P.O.\ Box 19395-5746, Tehran, Iran
\and       School of Chemistry, The University of Nottingham, University Park,
           Nottingham, NG7 2RD, UK
\and       Royal Astronomical Society, Burlington House, Piccadilly, London
           W1J 0BQ, UK}
\date{Submitted: 2015}
\abstract
{The Sun traverses a low-density, hot entity called the Local Bubble. Despite
its relevance to life on Earth, the conditions in the Local Bubble and its
exact configuration are not very well known. Besides, there is some unknown
interstellar substance which causes a host of absorption bands across the
optical spectrum, called Diffuse Interstellar Bands (DIBs).}
{We have started a project to chart the Local Bubble in a novel way, and learn
more about the carriers of the DIBs, by using DIBs as tracers of diffuse gas
and environmental conditions.}
{We have conducted a high signal-to-noise spectroscopic survey of 670 nearby
early-type stars, to map DIB absorption in and around the Local Bubble. The
project started with a Southern hemisphere survey conducted at the European
Southern Observatory's New Technology Telescope and has since been extended to
an all-sky survey using the Isaac Newton Telescope.}
{In this first paper in the series, we introduce the overall project and
present the results from the Southern hemisphere survey. We make available a
catalogue of equivalent-width measurements of the DIBs at 5780, 5797, 5850,
6196, 6203, 6270, 6283 and 6614 \AA, the interstellar Na\,{\sc i} D lines at
5890 and 5896 \AA, and the stellar He\,{\sc i} line at 5876 \AA. We find that
the 5780 \AA\ DIB is relatively strong throughout, as compared to the 5797
\AA\ DIB, but especially within the Local Bubble and at the interface with
more neutral medium. The 6203 \AA\ DIB shows a similar behaviour, but with
respect to the 6196 \AA\ DIB. Some nearby stars show surprisingly strong DIBs
whereas some distant stars show very weak DIBs, indicating small-scale
structure within as well as outside the Local Bubble. The sight-lines with
non-detections trace the extent of the Local Bubble especially clearly, and
show it opening out into the Halo.}
{The Local Bubble has a wall which is in contact with hot gas and/or a harsh
interstellar radiation field. That wall is perforated though, causing leakage
of radiation and possibly hot gas. On the other hand, compact self-shielded
cloudlets are present much closer to the Sun, probably within the Local Bubble
itself. As for the carriers of the DIBs, our observations confirm the notion
that these are large molecules, whose differences in behaviour are mainly
governed by their differing resilience and/or electrical charge, with more
subtle differences possibly related to varying excitation.}
\authorrunning{Bailey et al.}
\titlerunning{Probing the local diffuse ISM with DIBs. I.}
\maketitle

\section{Introduction}

The Solar System moves through a region of unusually low gas density -- the
Local Bubble -- initially discovered from the very low values of interstellar
attenuation of starlight within 100 pc of the Sun (Fitzgerald 1968; Lucke
1978; cf.\ Lallement et al.\ 2014a). Lallement et al.\ (2003) and Welsh et al.\
(2010) mapped the Local Bubble in Na\,{\sc i} and Ca\,{\sc ii} interstellar
absorption, revealing a central region of very low neutral/singly-ionized
atomic gas absorption out to a distance of about 80 pc in most Galactic
directions, and vertically for hundreds of parsecs -- presumably as far as the
Galactic Halo. However, the Ca\,{\sc ii} maps did not show any sharp increase
in absorption at a distance of 80 pc, leaving considerable uncertainty about
the interface (``wall'') between the tenuous local and denser surrounding
interstellar medium (ISM). It has also been surprisingly difficult to assess
the contents of the Local Bubble. Its possible origin in multiple supernova
explosions (Frisch 1981; Cox \& Anderson 1982; Ma\'{\i}z-Apell\'aniz 2001;
Breitschwerdt et al.\ 2009) seemed to be corroborated at first by the
detection of a diffuse soft-X-ray background (Snowden et al.\ 1998). While the
putative million-degree gas was not seen in the expected extreme-UV emission
by the CHIPS satellite (Hurwitz, Sasseen \& Sirk 2005) nor in O\,{\sc vi} 1032
\AA\ absorption (Cox 2005; Barstow et al.\ 2010), recent works making use of
soft X-ray shadows, Planck mm-wavelength data and Local Bubble maps have now
considerably reinforced the existence of such hot gas within the cavity
(Snowden et al.\ 2015). The picture is complicated by the interaction between
the heliosphere and surrounding ISM, such as charge exchange between the solar
wind ions and interstellar atoms (e.g., Lallement et al.\ 2014b; Zirnstein,
Heerikhuisen \& McComas 2015; but see Galeazzi et al.\ 2014), indicating local
``fluff'' of a more neutral kind (Frisch 1998). Any novel way or tracer that
can be used to map the Local Bubble is therefore highly desirable.

Bring on the Diffuse Interstellar Bands (DIBs). These are a set of over 400
broad optical absorption features which are ubiquitous in the ISM (Herbig
1995; Sarre 2006). The nature of their carriers has been much debated since
Mary Lea Heger first observed them almost a century ago (Heger 1922); the
current consensus is that they are large carbonaceous molecules, with some
resistance to UV radiation (Tielens 2014). Observations of DIBs have shown
that they are sensitive to their environment; for instance, the ratio of the
5780 and 5797 \AA\ DIBs is diagnostic of the level of irradiation and/or
kinetic temperature (Cami et al.\ 1997). Differences in absorption between
close binary stars have been detected, indicating DIBs trace also tiny-scale
structure in the ISM (Cordiner et al.\ 2013).

In this project we use DIBs to map the Local Bubble, by observing hundreds of
early-type stars within a few hundred pc of the Sun. These stars have well
determined distances and proper motions and future observations could reveal
tiny-scale structure. Also, by probing very local interstellar matter the
sight-lines suffer less from overlapping cloud structures and hence the
behaviour of the various tracers may be simpler to interpret and more telling.
DIBs have been used successfully to map extra-planar gas (van Loon et al.\
2009; Baron et al.\ 2015) and the Magellanic Clouds (van Loon et al.\ 2013;
Bailey et al.\ 2015) as well as parts of the Milky Way (Kos et al.\ 2014;
Zasowski et al.\ 2015). The elusiveness of the gas within the Local Bubble,
however, requires very low noise levels on the measurements and thus calls for
dedicated observations. We have initiated such a dedicated observing campaign,
first covering the Southern hemisphere presented here, subsequently extending
it to cover also the Northern hemisphere (Farhang et al.\ 2015a,b -- Papers II
and III). The combined, all-sky survey is used to apply inversion techniques
to reconstruct a complete three-dimensional map of the density of the DIB
carriers (Farhang et al., in prep.\ -- Paper IV). In a fifth paper in the
series we examine time variations of interstellar absorption due to the proper
motions of the stars, incorporating measurements for stars in common with
previous works (e.g., Raimond et al.\ 2012; Puspitarini, Lallement \& Chen
2013).

\section{Measurements -- Southern hemisphere survey}

\subsection{Targets}

%
\begin{table*}
\caption{Catalogue of observed stars. Only the first three rows are displayed;
the full catalogue of 238 objects is made available at CDS. The columns list
the HD number, equatorial coordinates $\alpha$ and $\delta$ and Galactic
coordinates $l$ and $b$, {\it Hipparcos} proper motions $\mu$ and parallax
$\pi$ (van Leeuwen 2007) and distance from the Sun $d$, V-band magnitude and
spectral type (from Simbad -- Wenger et al.\ 2000).}
\begin{tabular}{lcccccccccl}
\hline\hline
HD                                  &
\multicolumn{2}{c}{$\alpha$\ \ \ \ \, (J2000)\ \ \ \ \, $\delta$}  &
$l$                                 &
$b$                                 &
$\mu_\alpha$                         &
$\mu_\delta$                         &
$\pi$                               &
$d$                                 &
$V$                                 &
type                                \\
                                    &
($^{\rm h}$ $^{\rm m}$ $^{\rm s}$)     &
($^\circ$ $^\prime$ $^{\prime\prime}$) &
($^\circ$)                           &
($^\circ$)                           &
\multicolumn{2}{c}{(mas yr$^{-1}$)}  &
(mas)                               &
(pc)                                &
(mag)                               &
                                    \\
\hline
       955                          &
00 13 55.60                         &
$-$17 32 43.3                       &
78.8928                             &
$-77.0855$                          &
$17.11\pm1.05$                      &
$4.74\pm0.45$                       &
$3.16\pm0.71$                       &
\llap{3}16                          &
7.37                                &
B4V                                 \\
\llap{2}913                         &
00 32 23.78                         &
+06 57 19.7                         &
\llap{1}14.5465                     &
$-55.6060$                          &
$44.75\pm0.47$                      &
$2.33\pm0.35$                       &
\llap{1}$2.35\pm0.55$               &
81\rlap{.0}                         &
5.70                                &
B9.5V                               \\
\llap{3}580                         &
00 38 31.85                         &
$-$20 17 47.7                       &
98.8792                             &
$-82.5581$                          &
$22.24\pm0.38$                      &
$7.68\pm0.25$                       &
$3.97\pm0.39$                       &
\llap{2}52                          &
6.71                                &
B8V                                 \\
\hline
\end{tabular}
\end{table*}

The targets were chosen from the Na\,{\sc i}\,D Local Bubble survey of
Lallement et al.\ (2003). The criteria used to select the targets were that
they should be bright stars with B-band magnitudes $<8.7$, be of spectral type
earlier than A5 (to provide a clean continuum), have well-known distances from
{\it Hipparcos} and accurately known, high proper motions (for any future
time-variability studies of the tiny scale structure of the ISM). In all, 238
stars were observed, with V-band magnitudes ranging from 2.9 to 8.4. Of these,
109 are of early-A spectral type, 128 of spectral type B, and one of spectral
type O9.5. Information on the observed stars is listed in Table 1, the full
version of which is available electronically at the Centre de Donn\'ees
astronomiques de Strasbourg (CDS).

The locations on the sky are shown in Figure 1. There are some regions which
have not been sampled as uniformly as others; this is partly due to the
distribution of potential targets -- with most of them being concentrated
towards the Galactic plane -- and partly due to adverse observing conditions.
To gain a better appreciation of the sampling, we show the distributions of
nearest-neighbour distance in 3D (in pc) in Figure 2; typical sampling of 3D
space is on 10--40 pc scales.

%
\begin{figure}
\centerline{\epsfig{figure=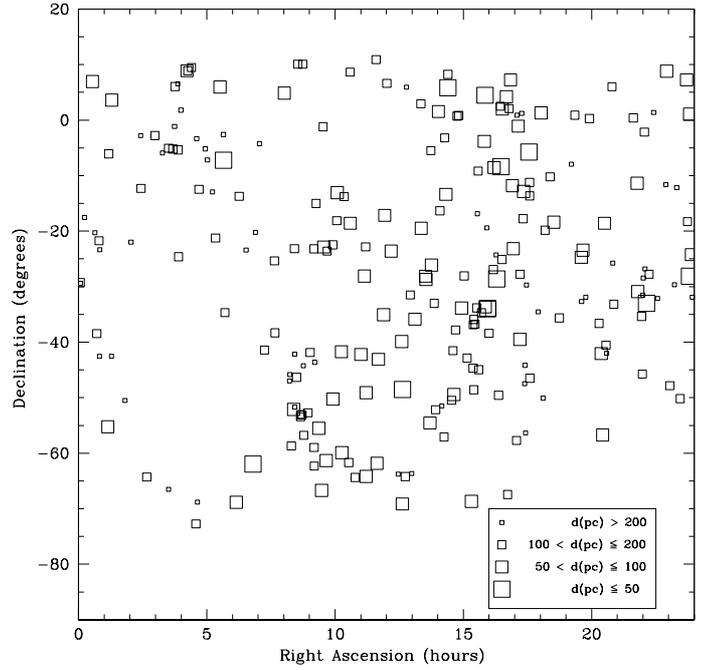,width=90mm}} 
\caption[]{Distribution on the sky of the 238 observed stars as part of the
Southern hemisphere DIBs survey. The size of the symbol indicates the distance
from the Sun.}
\end{figure}

%
\begin{figure}
\centerline{\epsfig{figure=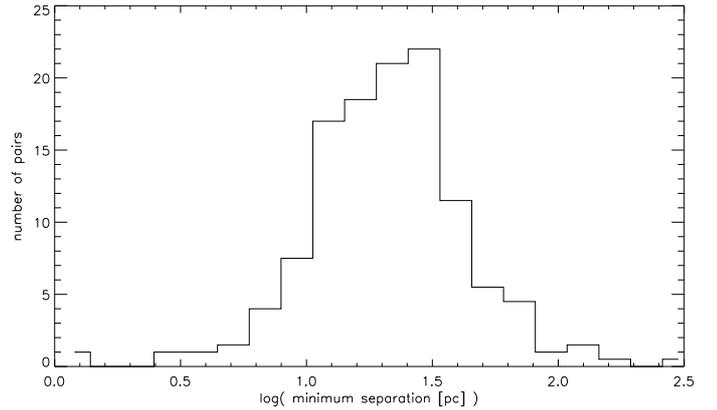,width=90mm}} 
\caption{Distribution of nearest-neighbour distance, in pc, in the Southern
hemisphere DIBs survey.}
\end{figure}

\subsection{Observations}

We obtained long-slit spectroscopic data using the 3.5m New Technology
Telescope (NTT) at the European Southern Observatory, La Silla, Chile. The
observations were performed in visitor mode on the nights of
$21^{\rm st}$--$23^{\rm rd}$ March 2011 and $15^{\rm th}$--$17^{\rm th}$ August
2011 (ESO programme 087.C-0405(A+B); PI: M.Bailey) and
$19^{\rm th}$--$22^{\rm nd}$ August 2012 (ESO programme 0.89.C-0797; PI:
M.Bailey).

The Faint Object Spectrograph and Camera (EFOSC2) were used with grism \#20
and the $0\rlap{.}^{\prime \prime}3$ blue offset slit; this gave a resolving
power of $R=\lambda/\Delta\lambda=5500$, resulting in a two-pixel sampled
spectral resolution of $\approx 1.1$ \AA\ (no binning of pixels on the CCD).
This combination of grism and offset slit covers the wavelength range
5672--6772 \AA, which includes the major $\lambda$5780, 5797, 5850, 6196,
6203, 6269, 6283 and 6614 DIBs. Grism \#20 introduces a lateral shift of the
beam. Therefore, the target was placed at column 1680 during acquisition to
ensure the photons arrived approximately at column 1107 once the grism was put
in place, which is an area on CCD\#40 that is free from bad pixels (the
dispersion direction runs along the columns of the CCD).

Exposure times were determined with the goal of reaching maximum count levels
of $\sim40,000$--$50,000$ counts per pixel, which is close to but below the
CCD saturation level ($65,535$ counts); these ranged from 3.4 seconds to 847
seconds. The CCD was read out in its fast mode, as the read-out-noise was
negligible compared to the photon noise. A minimum of three exposures were
taken for each target to achieve a total signal-to-noise level (S/N) of
$\sim2,000$ after the spectra had been combined. If the maximum counts per
pixel was $<20,000$ a further three exposures were made. Exposure times were
altered to accommodate changing conditions -- the seeing varied between
$0\rlap{.}^{\prime \prime}4$ and $2\rlap{.}^{\prime \prime}4$; some exposures
resulted in saturation of the detector during moments of excellent seeing as
the centering on the narrow slit became problematic.

Before the start of each night a set of calibration frames were taken. These
comprised a large number of bias frames ($>30$), to remove the electronic
offset level; dome flat fields ($\sim60$, using $4\times1,000$ W
quartz--halogen lamps) to correct for pixel-to-pixel sensitivity and
throughput variations across the CCD; and He+Ar arc lamp exposures for
wavelength calibration. The large number of bias and flat field frames was
necessary in order for these not to limit the S/N on the calibrated spectra.


\subsection{Data processing}

Standard data reduction steps were applied: creation of master frames
representing the bias level and flat field, respectively; subtraction of the
master bias frame from the master flat and science frames; division of the
science frames by the master flat; wavelength calibration on the basis of the
He+Ar arc lamp spectrum. The spectra were then extracted from the calibrated
2D frames using the optimal (weighted) extraction method of Horne (1986). This
also removed some particularly bright cosmic ray impacts upon the detector --
though these were few in number due to the generally short integration times.

Two issues were encountered during data processing: bias instability and
wavelength calibration offsets (for more details see Bailey 2014). The fast
read-out mode makes use of two amplifiers, resulting in different bias levels
for both halves of the detector. The bias level was found to vary over the
course of the measurements in an unpredictable manner. It was thus decided in
the second and third of the three observing runs to also obtain a bias frame
after every science observation. The offset slit results in part of the CCD
remaining unilluminated and we therefore decided to use the recorded count
level at the edge furthest away from the illuminated part of the CCD, to
determine an offset value with which to correct the master bias
frame\footnote{A gradient of intensity was observed indicating light still
somehow made it to the ``unilluminated'' part of the CCD.}.

EFOSC2 was originally designed to be mounted at the Cassegrain focus (of the
ESO 3.6m telescope) where the aperture wheel moved in a horizontal plane
inclining only with the telescope. Now that EFOSC2 is located at the Nasmyth
focus (of the NTT), flexure problems are experienced. This is due to the fact
that the instrument is now turned on its side and gravity can influence the
position of the slits depending on the rotator position of the instrument. The
effect of the flexure is a variation of the position of the slit on the
detector in the dispersion direction, which results in shifts of the
wavelength scale that can be as big as several \AA\ (EFOSC2 User's Manual). We
corrected for these offsets by recalibrating each spectrum using the telluric
O$_2$ absorption line at $\lambda=6278.1$ \AA.

\subsection{Methodology}

We here first describe the analysis of the 5780 \AA\ DIB. A shallow broad
absorption as well as a stellar absorption line at 5785 \AA\ were fitted with
a second-order polynomial and a Gaussian function, respectively, at the same
time as fitting a Gaussian function to the 5780 \AA\ DIB. While we considered
fitting a Lorentzian instead, the above strategy worked well and is more in
line with the choice of Gaussian found in the literature (Herbig 1975, Hobbs
et al.\ 2008; Jenniskens \& D\'esert 1994; Kre{\l}owski \& Westerlund 1988;
Westerlund \& Kre{\l}owski 1988). The wavelength and typical width of the DIB
was taken from the DIB catalogue of Jenniskens \& D\'esert (1994), but these
were allowed to vary by $\pm2$ \AA\ and 0.5 and $2\times$, respectively
($\pm2$ \AA\ and $\pm0.2$ \AA\ for the stellar line). The instrumental
resolution of 1.1 \AA\ was sufficiently less than the expected width of the
5780 DIB ($\sim2$ \AA) to determine the intrinsic profile shape, but this was
not the case for the other DIBs under consideration here ($\sim0.6$--1.2 \AA)
except for the 6270 \AA\ DIB ($\sim3$--4 \AA). The actual fitting was coded in
{\sc idl} using the {\sc mpfit} routines, in the range 5773--5788 \AA.

The equivalent width $EW$ was determined by integrating the Gaussian function:
\begin{equation}
EW = \int_{-\infty}^\infty
  A \exp\left(-\ \frac{(\lambda-\lambda_0)^2}{2\sigma^2}\right)
= A \sigma \sqrt{2\pi},
\end{equation}
where $A$ is the peak intensity, $\lambda_0$ the central wavelength and
$\sigma$ the width (with the Full Width at Half Maximum, $FWHM=2.355\sigma$).
The errors $e_0$ resulting from {\sc mpfitexpr} were scaled according to the
$\chi^2$ deviations from the fit and the degrees of freedom $f$ (the number of
spectral elements minus the number of fitting parameters) as in:
\begin{equation}
e = e_0 \sqrt{\chi^2/f}
\end{equation}
Hence the error in the equivalent width is obtained from:
\begin{equation}
e[EW] = \sqrt{2\pi \left((e[A] \sigma)^2+(A e[\sigma])^2\right)}
\end{equation}

%
\begin{table}
\caption{Overview of the Gaussian fitting constraints, in \AA.}
\begin{tabular}{llclc}
\hline\hline
Feature                &
\multicolumn{2}{c}{\llap{---}--------- $\lambda_0$ ---------\rlap{---}} &
\multicolumn{2}{c}{\llap{---}--------- $\sigma$ ---------\rlap{---}}    \\
                       &
guess                  &
range                  &
guess                  &
range                  \\
\hline
5780 DIB               &
5780.6                 &
$\pm2.00$              &
1.25                   &
0.625--2.500           \\
5797 DIB               &
$\lambda_{5780}$+\ \ 16.50 &
$\pm0.50$              &
1.00                   &
0.500--2.000           \\
5850 DIB               &
$\lambda_{5780}$+\ \ 69.60 &
$\pm0.50$              &
1.00                   &
0.500--2.000           \\
He\,{\sc i}            &
5875.6                 &
$\pm2.60$              &
1.30                   &
0.650--2.600           \\
Na\,{\sc i} D$_2$      &
5889.95                &
$\pm2.00$              &
1.00                   &
0.500--2.000           \\
Na\,{\sc i} D$_1$      &
$\lambda_{\rm D2}$\ \ +\ \ \ \ 5.97 &
0                      &
$\sigma_{\rm D2}$       &
0                      \\
6196 DIB               &
$\lambda_{5780}$+415.60 &
$\pm0.50$              &
0.70                   &
0.350--1.400           \\
6203 DIB               &
$\lambda_{5780}$+422.60 &
$\pm0.75$              &
1.00                   &
0.500-2.000            \\
6270 DIB               &
$\lambda_{5780}$+489.50 &
$\pm0.50$              &
1.00                   &
0.500-2.000            \\
6283 DIB               &
$\lambda_{5780}$+503.70 &
$\pm0.50$              &
1.00                   &
0.500-2.000            \\
6614 DIB               &
$\lambda_{5780}$+833.10 &
$\pm0.80$              &
1.00                   &
0.500-2.000            \\
\hline
\end{tabular}
\end{table}

The constraints on the fitting to the other absorption lines and DIBs are
summarised in Table 2. In particular, the wavelengths of the other DIBs were
guided by that of the strongest, 5780 \AA\ DIB and allowed to vary much less
(typically $\pm0.5$ \AA) to avoid spurious results. Likewise, the wavelength
of the D$_1$ (red) component of the Na\,{\sc i} doublet was tied to that of
the stronger D$_2$ (blue) component; both components were fitted
simultaneously. The spectral baseline was fitted with a first-order polynomial
rather than the second-order polynomial above, except in some cases such as
the Na\,{\sc i} doublet where a shallow, broad absorption component was
accounted for with a second-order polynomial -- a second fit was obtained
using a first-order polynomial, which was divided by the initial fit
(including the second-order polynomial and two Gaussian functions) to reveal
the broad absorption component. The 6196 and 6203 \AA\ DIBs were also fitted
simultaneously and here too a second-order polynomial was required for the
continuum fitting. A second-order polynomial was also required for the 6270
\AA\ DIB continuum as well as to take account of the telluric O$_2$ line near
the 6283 \AA\ DIB (before fitting the DIB with a Gaussian and first-order
polynomial). Finally, a second-order polynomial was required to account for
modest fringing near the reddest of the DIBs under consideration, at 6614 \AA.

%
\begin{table}
\caption{Spectral line measurements. $EW$ stands for equivalent width, with
$e[EW]$ its uncertainty.
Flags $f$ are 1 or 0 for a detection or non-detection, respectively; $f=2$
means dubious (excluded from further analysis). The full, transposed version
of this table is available on CDS.}
\begin{tabular}{llll}
\hline\hline
HD                     & 955    & 2913   & 3580   \\
\hline
$EW_{5780}$ (\AA)       & 0.008  & 0.016  & 0.020  \\
$e[EW]_{5780}$ (\AA)    & 0.0020 & 0.0030 & 0.0039 \\
$(EW/e[EW])_{5780}$     & 4.1    & 5.4    & 5.2    \\
$EW_{5797}$ (\AA)       & -      & -      & 0.008  \\
$e[EW]_{5797}$ (\AA)    & -      & -      & 0.0170 \\
$(EW/e[EW])_{5797}$     & -      & -      & 0.5    \\
$EW_{5850}$ (\AA)       & 0.006  & 0.011  & -      \\
$e[EW]_{5850}$ (\AA)    & 0.0094 & 0.0038 & -      \\
$(EW/e[EW])_{5850}$     & 0.7    & 2.9    & -      \\
$EW_{\rm He}$ (\AA)     & 0.621  & 0.058  & 0.015  \\
$e[EW]_{\rm He}$ (\AA)  & 0.0180 & 0.0061 & 0.0104 \\
$(EW/e[EW])_{\rm He}$   & 34.5   &  9.6   & 1.4    \\
$EW_{\rm NaD2}$ (\AA)    & 0.067  & 0.097  & 0.022  \\
$e[EW]_{\rm NaD2}$ (\AA) & 0.0078 & 0.0148 & 0.0106 \\
$EW_{\rm NaD1}$ (\AA)    & 0.010  & 0.043  & -      \\
$e[EW]_{\rm NaD1}$ (\AA) & 0.0046 & 0.0109 & -      \\
$(EW/e[EW])_{\rm NaD2}$  & 8.6    & 6.5    & 2.1    \\
$(EW/e[EW])_{\rm NaD1}$  & 2.1    & 3.9    & -      \\
$EW_{6196}$ (\AA)       & 0.003  & -      & 0.008  \\
$e[EW]_{6196}$ (\AA)    & 0.0021 & -      & 0.0026 \\
$EW_{6203}$ (\AA)       & -      & -      & -      \\
$e[EW]_{6203}$ (\AA)    & -      & -      & -      \\
$(EW/e[EW])_{6196}$     & 1.5    & -      & 3.1    \\
$(EW/e[EW])_{6203}$     & -      & -      & -      \\
$EW_{6270}$ (\AA)       & 0.015  & 0.002  & -      \\
$e[EW]_{6270}$ (\AA)    & 0.0085 & 0.0035 & -      \\
$(EW/e[EW])_{6270}$     & 1.7    & 0.5    & -      \\
$EW_{6283}$ (\AA)       & -      & -      & -      \\
$e[EW]_{6283}$ (\AA)    & -      & -      & -      \\
$(EW/e[EW])_{6283}$     & -      & -      & -      \\
$EW_{6614}$ (\AA)       & 0.007  & -      & 0.008  \\
$e[EW]_{6614}$ (\AA)    & 0.0066 & -      & 0.0044 \\
$(EW/e[EW])_{6614}$     & 1.1    & -      & 1.9    \\
$f_{5780}$              & 1      & 0      & 1      \\
$f_{5797}$              & 0      & 0      & 0      \\
$f_{5850}$              & 0      & 0      & 0      \\
$f_{\rm He}$            & 1       & 1     & 0      \\
$f_{\rm NaD2}$           & 1      & 1      & 0      \\
$f_{\rm NaD1}$           & 0      & 2      & 0      \\
$f_{\rm Na\ dip}$        & 0      & 1      & 0      \\
$f_{6196}$              & 0      & 0      & 0      \\
$f_{6203}$              & 0      & 0      & 0      \\
$f_{6270}$              & 0      & 0      & 0      \\
$f_{6283}$              & 0      & 0      & 0      \\
$f_{6614}$              & 0      & 0      & 0      \\
\hline
\end{tabular}
\end{table}

All spectra and fits were vetted by eye, guided by the value for $EW/e[EW]$,
and only fits deemed reliable were retained. The full table of measurements is
available at CDS, with the first three entries shown in Table 3.

\section{Results}

\subsection{Detected spectral lines and bands}

%

The He\,{\sc i} (D$_3$) line is stellar in origin but the Na\,{\sc i} D$_1$ \&
D$_2$ doublet is mostly interstellar. The latter sits on a broader shallow
absorption component, which is a combination of several telluric water lines
typically one per cent of the continuum level (see Bailey 2014 for more
details). The weak Na\,{\sc i} absorption, $\sim5$\% of the continuum on
average, illustrates how challenging it is to map the ISM within $\sim100$ pc
around the Sun.

%

The DIBs are more challenging still, less than one per cent of the continuum.
The 5780 \AA\ DIB is $\sim0.7$\% deep on average, but clearly detected in many
of our targets. The nearby and diagnostic comparison 5797 \AA\ DIB is $<0.2$\%
below the continuum. The 5850 \AA\ DIB is very feeble and being next to a much
stronger stellar feature to the red it is omitted from further analysis. The
6196 \& 6203 \AA\ DIBs are only $\sim0.1$\% deep on average, but sharp and
more isolated and thus more useful. The 6270 \AA\ DIB has a depth of typically
a few 0.1\% and is safe from the telluric O$_2$ absorption to red, but it is
broader and difficult to detect in individual targets and hence it is omitted
from further analysis. Likewise, the 6283 \AA\ DIB is omitted; it is but a
mere fraction of a per cent of continuum and sits on the wing of the strong
telluric O$_2$ absorption to the blue. The 6614 \AA\ DIB on the other hand is
only 0.2\% on average but it is well detected often enough to be useful for
analysis.


\subsection{Correlations}

%
\begin{figure*}
\centerline{\vbox{
\hbox{
\epsfig{figure=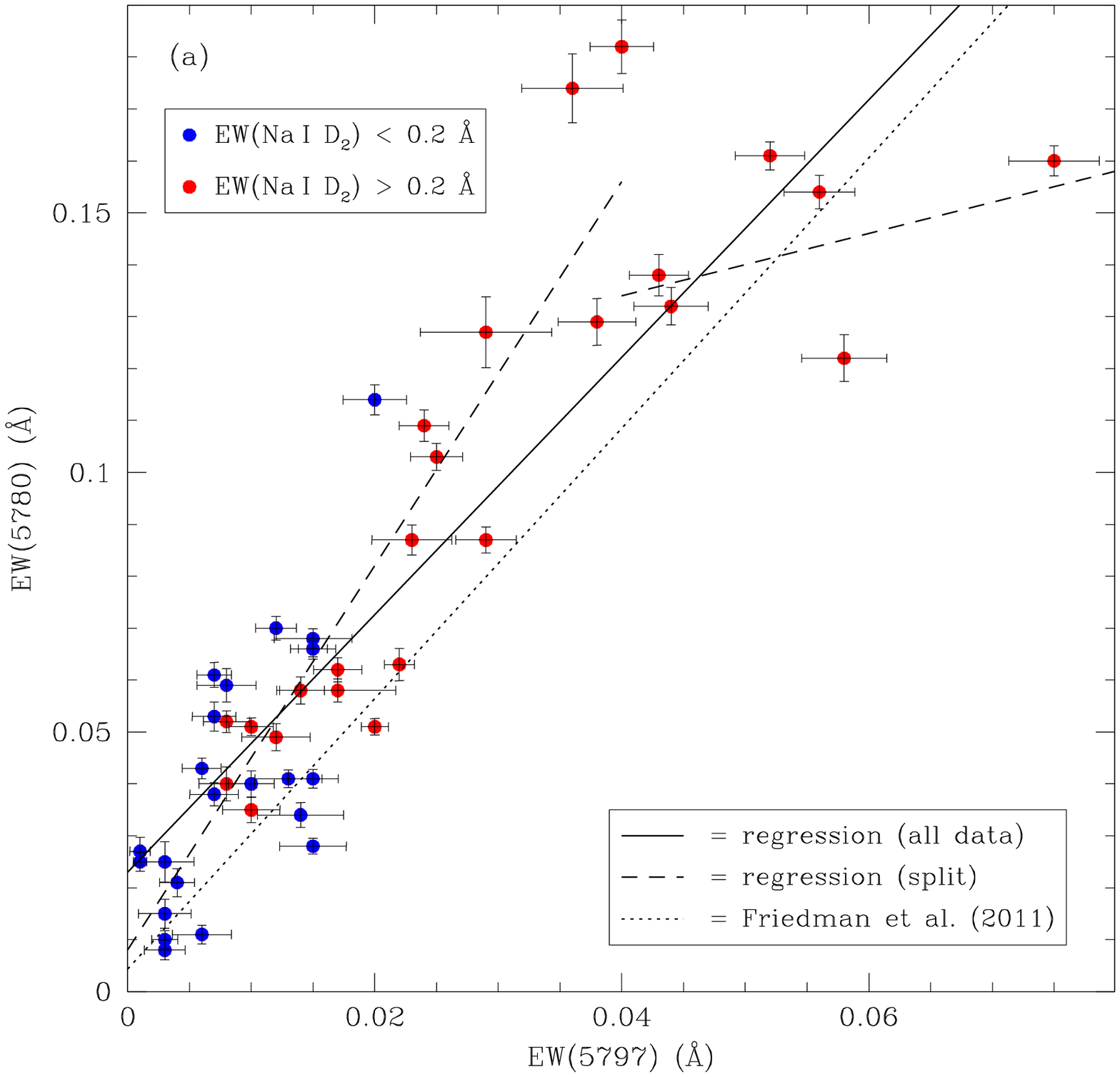,width=60mm}
\epsfig{figure=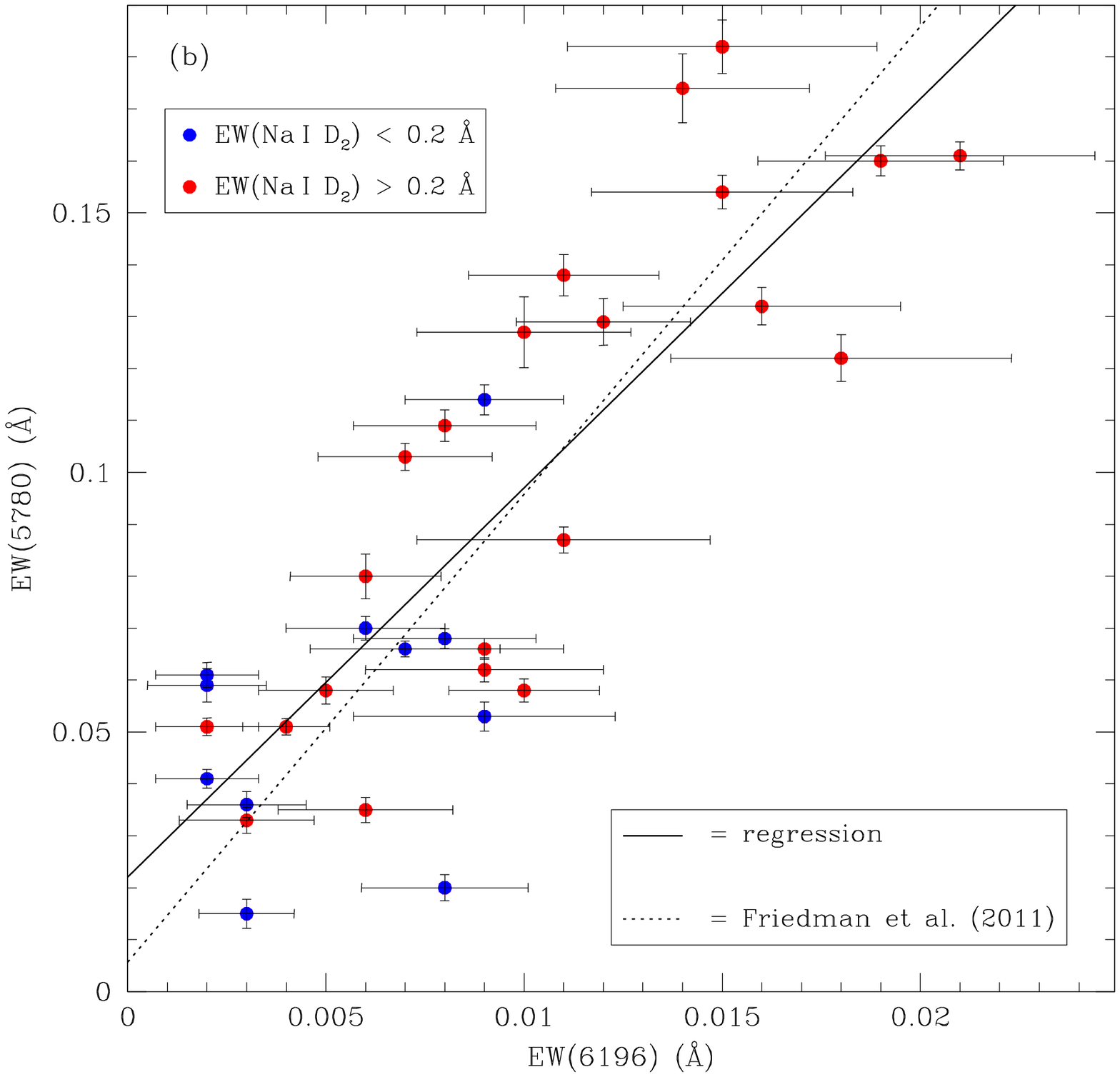,width=60mm}
\epsfig{figure=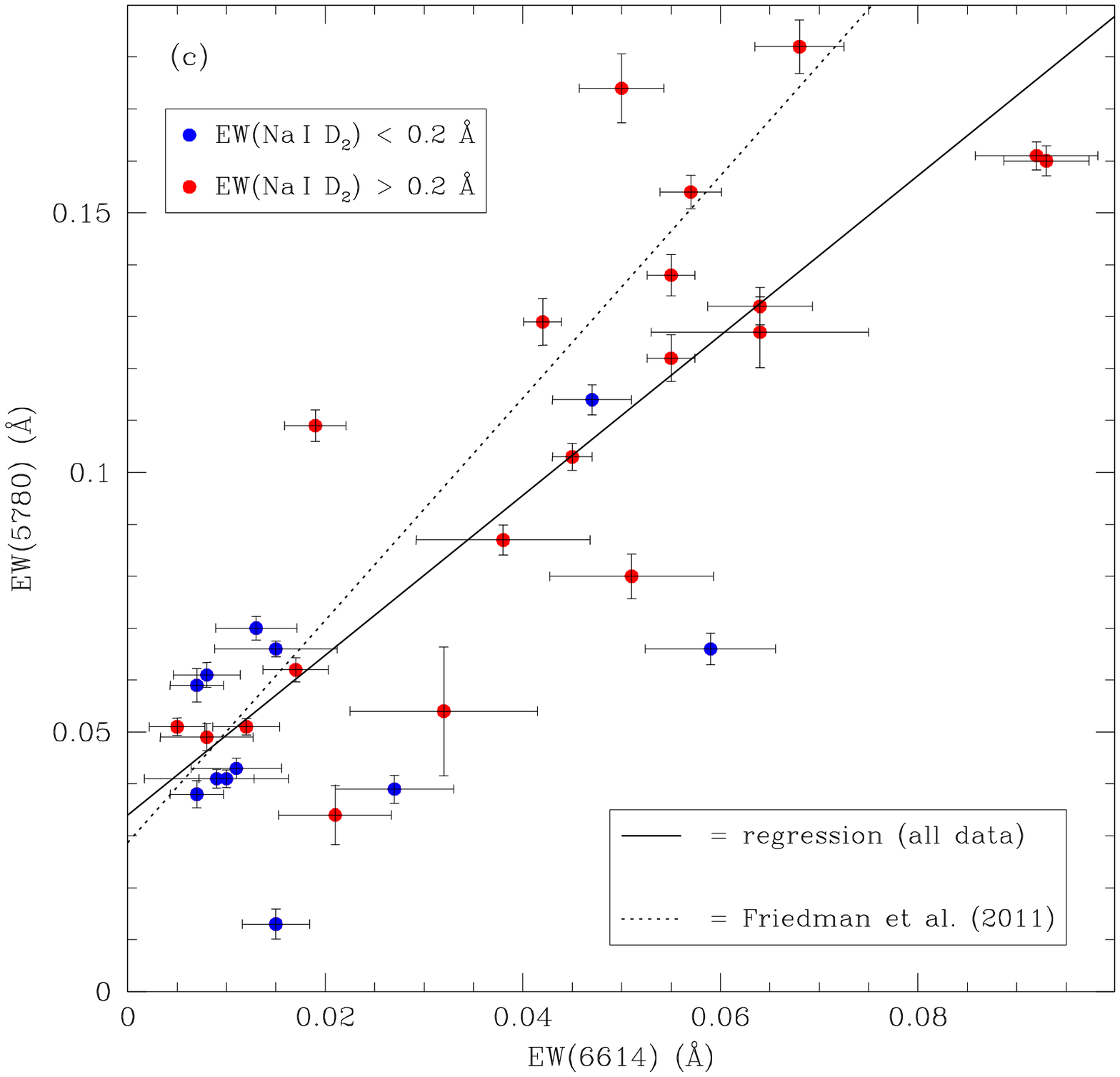,width=60mm}}
\hbox{
\epsfig{figure=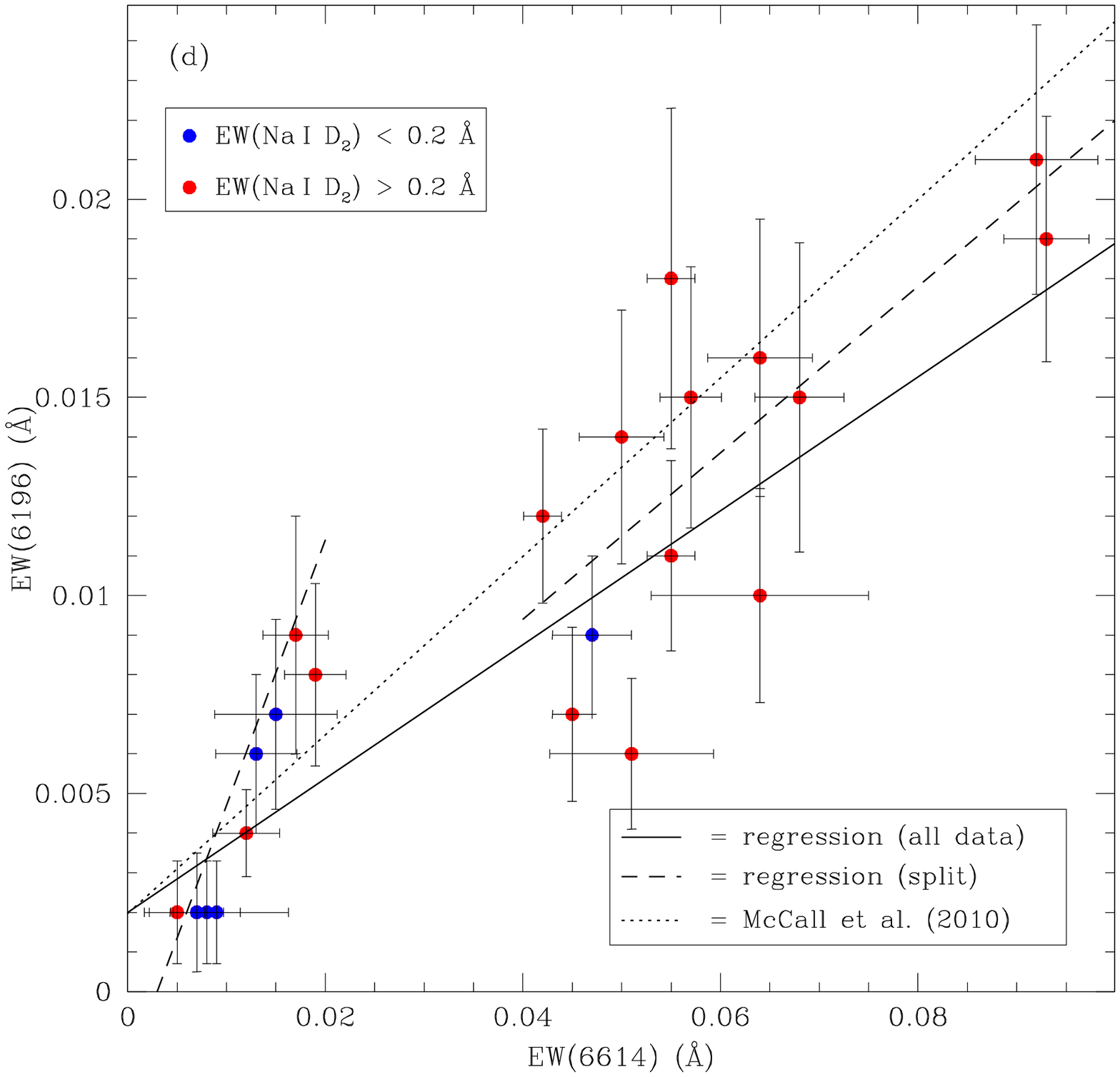,width=60mm}
\epsfig{figure=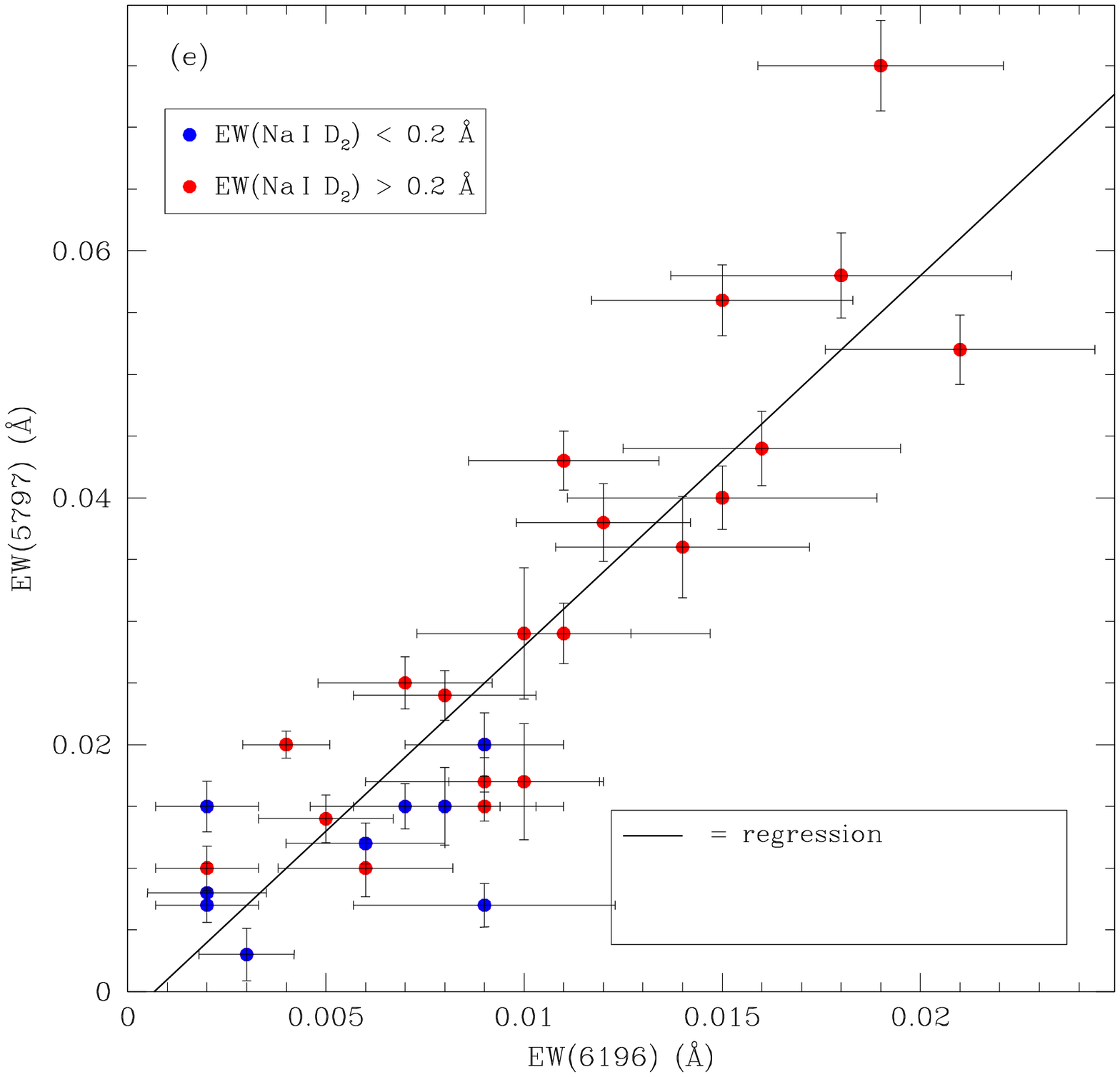,width=60mm}
\epsfig{figure=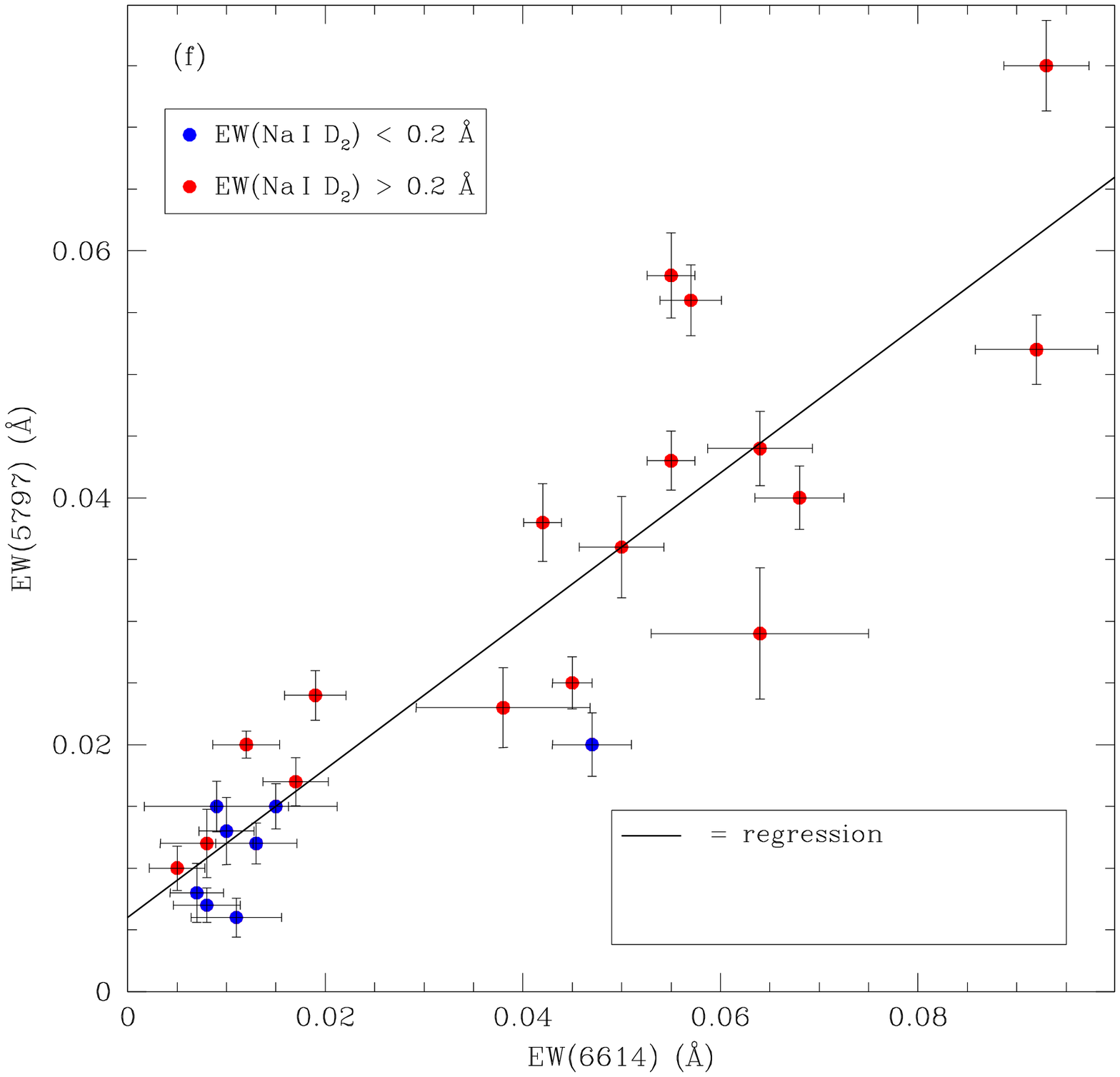,width=60mm}
}}}
\caption[]{Correlations between the 5780, 5797, 6196 and 6614 \AA\ DIBs.
Distinction is made on the basis of the Na\,{\sc i} absorption. Linear
regression fits are overplotted, including some from the literature.}
\end{figure*}

%
\begin{table*}
\caption{Linear regression analysis of DIB--DIB correlations of the form
$Y=a+bX$ with correlation coefficient $r$. Where the Student $t$ statistic is
given, the critical value of $t$ is for a 1\% chance that the null hypothesis
is true.}
\begin{tabular}{clrcccccl}
\hline\hline
$Y/X$ pair            &
selection             &
$N$                   &
$a$ (\AA)             &
$b$                   &
$r$                   &
$t$                   &
$t_{\rm critical}$      &
verdict               \\
\hline
5780/5797             &
all                   &
46                    &
$0.023\pm0.005$       &
$2.48\ \ \pm0.20\ \ $ &
0.88                  &
\llap{1}2             &
2.7                   &
significant           \\
|                     &
$EW(5797)<0.04$       &
39                    &
$0.008\pm0.005$       &
$3.7\ \ \ \ \pm0.8\ \ \ \ $ &
0.90                  &
\llap{1}3             &
2.7                   &
significant           \\
|                     &
$EW(5797)\geq0.04$    &
7                     &
$0.11\ \ \pm0.04\ \ $ &
$0.6\ \ \ \ \pm0.6\ \ \ \ $ &
0.44                  &
1\rlap{.1}            &
4.0                   &
insignificant         \\
6196/6614             &
all                   &
22                    &
$0.002\pm0.000$       &
$0.189\pm0.022$       &
0.89                  &
8\rlap{.7}            &
2.8                   &
significant           \\
|                     &
$EW(6614)<0.02$       &
9                     &
\llap{$-$}$0.002\pm0.002$ &
$0.57\ \ \pm0.07\ \ $ &
0.95                  &
8\rlap{.0}            &
3.5                   &
significant           \\
|                     &
$EW(6614)>0.04$       &
13                    &
$0.001\pm0.001$       &
$0.21\ \ \pm0.06\ \ $ &
0.73                  &
3\rlap{.5}            &
3.1                   &
marginal              \\
5780/6196             &
all                   &
33                    &
$0.022\pm0.009$       &
$7.5\ \ \ \ \pm0.9\ \ \ \ $ &
0.82                  &
8\rlap{.0}            &
2.7                   &
significant           \\
5780/6614             &
all                   &
32                    &
$0.034\pm0.008$       &
$1.54\ \ \pm0.19\ \ $ &
0.83                  &
8\rlap{.2}            &
2.8                   &
significant           \\
5797/6196             &
all                   &
29                    &
\llap{$-$}$0.002\pm0.001$ &
$3.0\ \ \ \ \pm0.3\ \ \ \ $ &
0.89                  &
\llap{1}0             &
2.8                   &
significant           \\
5797/6614             &
all                   &
25                    &
$0.006\pm0.002$       &
$0.60\ \ \pm0.06\ \ $ &
0.89                  &
9\rlap{.4}            &
2.8                   &
significant           \\
\hline
\end{tabular}
\end{table*}

Figure 3a shows an excellent correlation between the equivalent widths of the
5780 and 5797 \AA\ DIBs, up to some point beyond which increasing strength of
the 5797 \AA\ DIB is not accompanied by a noticeable increase in the strength
of the 5780 \AA\ DIB. This occurs for $EW(5797)\geq0.04$ \AA, around
$EW(5780)\sim0.15$ \AA. Compared to the Friedman et al.\ (2011) regression fit
our sightlines seem to probe relatively strong 5780 \AA\ DIBs except for the
sight-lines with the strongest 5797 \AA\ DIB. The 5797 \AA\ DIB favours a
neutral environment, whereas the 5780 \AA\ DIB favours an ionized environment
so a strong detection of one should not necessarily be accompanied by a strong
detection of the other. In Table 4 the results are summarised from a linear
regression analysis and a Student $t$ test: the correlation is significant if
$t>t_{\rm critical}$, where
\begin{equation}
t=r\sqrt{\frac{N-2}{1-r^2}}
\end{equation}
with correlation coefficient $r$ and $N$ data points ($N-2$ degrees of freedom
in this case). These analyses have not taken into account the uncertainties of
the individual measurements. However, inspection of figure 3 shows that either
the errors are fairly uniform in the cases where the scatter does not vary
much along the correlation, or they increase with increasing equivalent widths
in concert with the scatter. The exact treatment of the uncertainties should
therefore not have a huge impact as they are already reflected in the scatter
among the values. Hence we conclude that a good correlation is seen when the
absorptions are fairly weak but the correlation breaks down above
$EW(5797)\geq0.04$ \AA.

As examples of variations in the $EW(5780)/EW(5797)$ ratio, figure 4 shows the
spectra of HD\,146284 and HD\,169009. They have a similar $EW(5780)$ but the
$EW(5797)=0.058$ \AA\ of HD\,169009 is twice the $EW(5797)=0.028$ \AA\ of
HD\,146284. While HD\,146284 is further away from the Sun ($d=315$ pc) than
HD\,169009 is ($d=102$ pc), HD\,146284 is also further away from the Galactic
Plane ($z=102$ pc) than HD\,169009 ($z=2.9$ pc): while the longer column
towards HD\,146284 through the more tenuous extra-planar medium results in
relatively strong diffuse band absorption, the harsher environment results in
a relatively weaker 5797 \AA\ DIB (cf.\ van Loon et al.\ 2009).

%
\begin{figure}
\centerline{\vbox{
\epsfig{figure=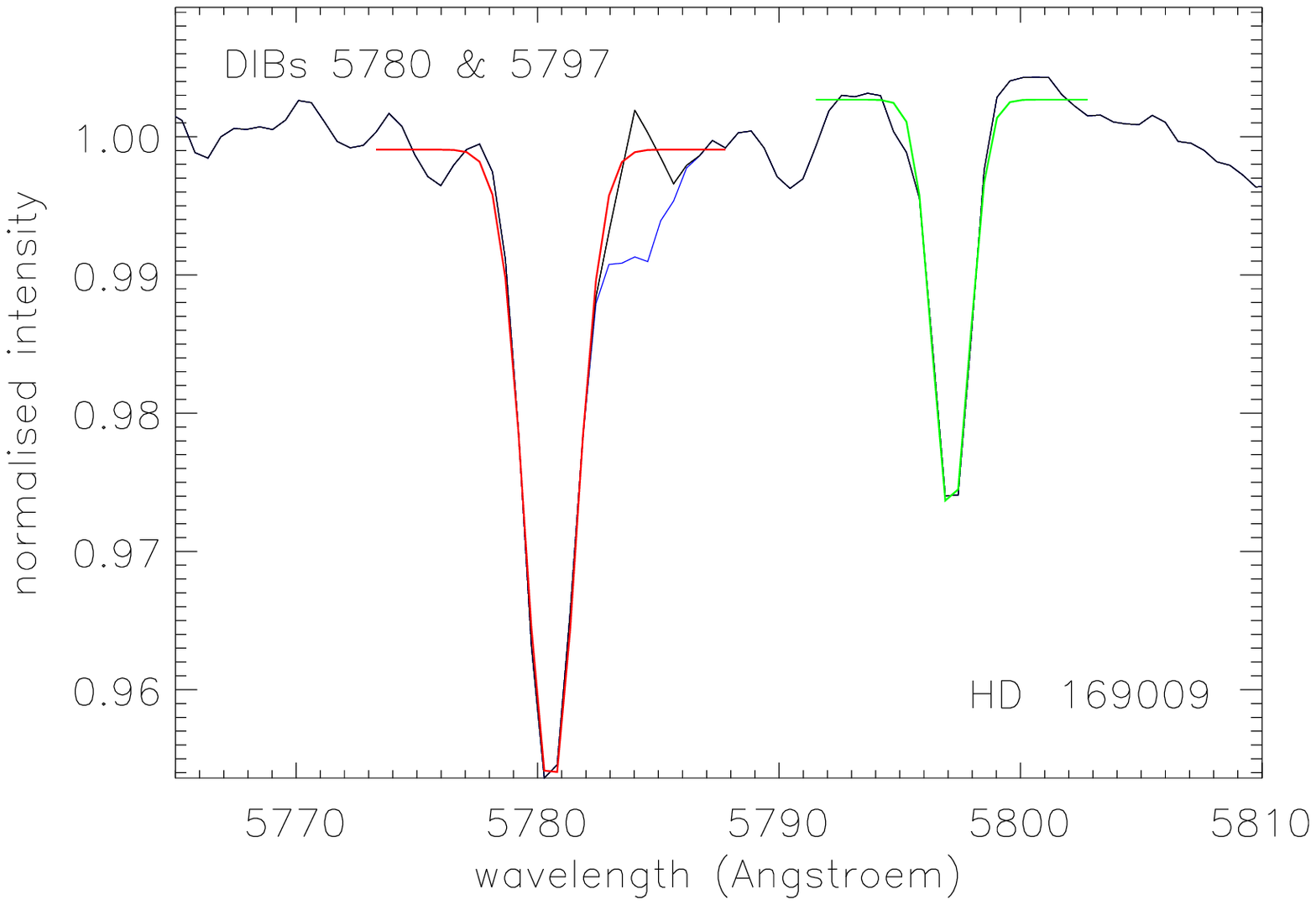,width=90mm} 
\epsfig{figure=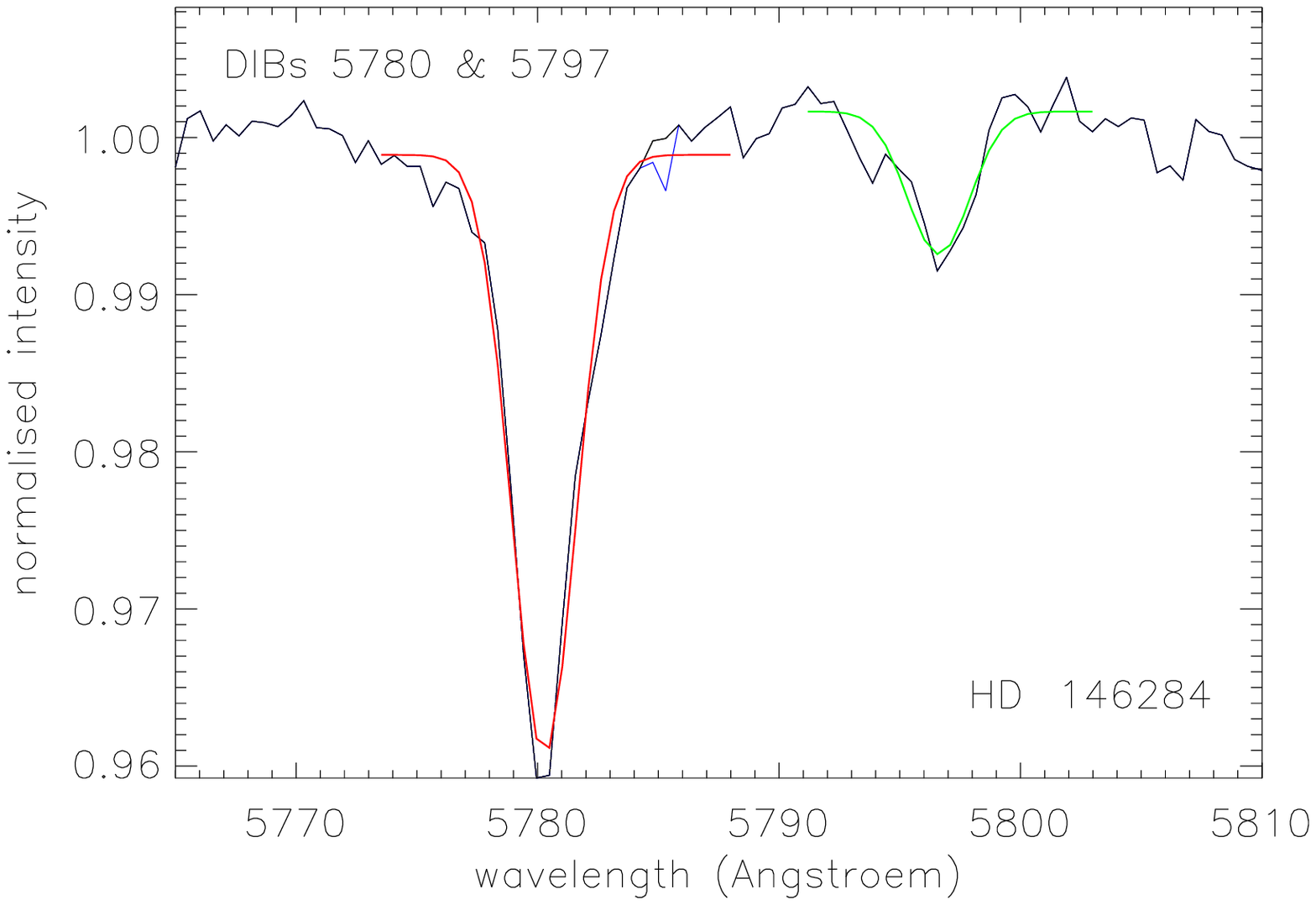,width=90mm} 
}}
\caption[]{The 5780 and 5797 \AA\ DIBs in the spectra of two stars at
different distance to the Sun ($d$) and Galactic Plane ($z$): {\it Top:}
HD\,169009 at $d,z=102,2.9$ pc; {\it Bottom:} HD\,146284 at $d,z=315,102$ pc.
The blue spectrum is the original spectrum before removing the stellar
absorption line around 5785 \AA; the red and green curves are the Gaussian
fits to the 5780 and 5797 \AA\ DIBs, respectively.}
\end{figure}

A good correlation is found between the 6196 and 6614 \AA\ DIBs (Fig.\ 3d).
This particular pair of DIBs is said to have a ``near perfect correlation''
(McCall et al.\ 2010). However, at the generally low $EW$ values in our study
there appears to be a change from a tight steep slope up to $EW(6196)<0.01$
\AA\ and $EW(6614)<0.02$ \AA, to a shallower relation at greater $EW$ values
(cf.\ Table 4). Considering a continuous, monotonic relation throughout, and
the regression fit from McCall et al., there is a suggestion in our data that
some sight-lines have an anomalously weak 6196 \AA\ DIB -- e.g., the datum at
$EW(6614)=0.051$, $EW(6196)=0.006$ \AA\ (HD\,145102). Oddly enough, the McCall
et al.\ data also showed some deviations at the weak end but in the opposite
sense, with the 6614 \AA\ DIB being weaker. This exemplifies the bias that can
be introduced when sampling different sight-lines, as well as distances (longer
columns may sample multiple clouds hence yielding smoother relationships).

%
%

The 5780 and 5797 \AA\ DIBs correlate well with the 6196 and 6614 \AA\ DIB
overall (Fig.\ 3b,c), but the correlations involving the 5797 \AA\ DIB are much
tighter than those involving the generally strong(er) 5780 \AA\ DIB (cf.\
Table 4). This exemplifies the different diagnostic power of the 5780 \AA\ DIB
as compared to the other three DIBs -- even though we noted that differences
between the 6196 and 6614 \AA\ DIB exist too (indeed, there is more scatter in
the relations between the 5780 and 5797 \AA\ DIBs with the 6614 \AA\ DIB than
with the 6196 \AA\ DIB). Compared to the regression fits from Friedman et al.\
(2011), the 5780 and 6196 \AA\ DIBs show an identical relation but the 6614
\AA\ DIB seems to be much stronger in certain (but not all) sight-lines; we
confirm the Friedman et al.\ suggestion that the 5780 \AA\ DIB is already seen
when the 6614 \AA\ DIB is not (a positive value for $a$ at more than
4-$\sigma$ significance). As an example, figure 5 shows two sight-lines where
the spectra have been superimposed for the regions around the 5780, 5797 and
6614 \AA\ DIBs; HD\,18650 has a larger Galactic latitude ($b=-25^\circ$) than
HD\,179406 ($b=-8^\circ$). Clearly, their $EW(5780)/EW(5797)$ ratio differs by
about a factor two, but the $EW(5797)/EW(6614)$ ratio is very similar.

%
\begin{figure}
\centerline{\epsfig{figure=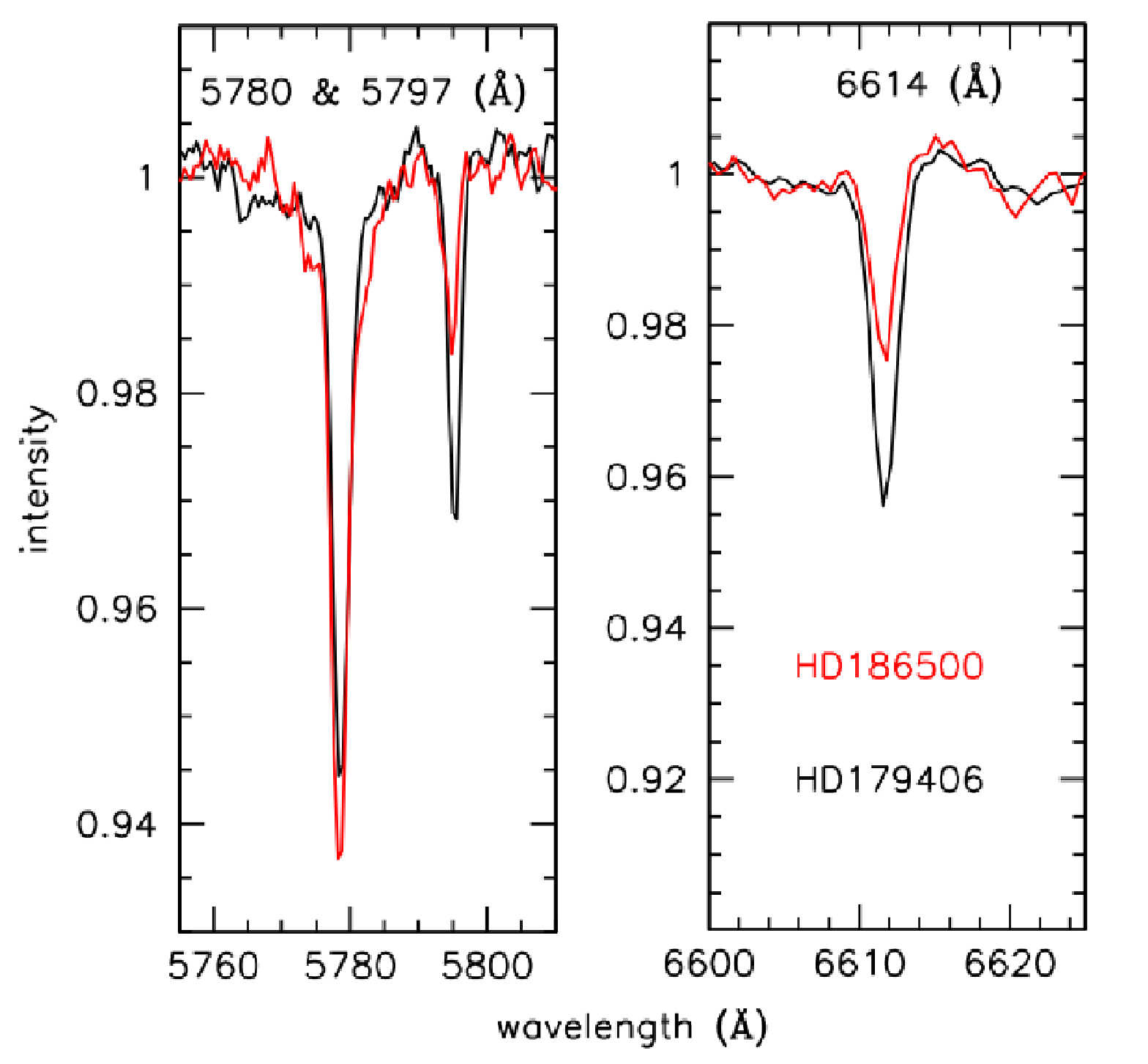,width=90mm}}
\caption[]{The 5780, 5797 and 6614 \AA\ DIBs in the spectra of HD\,186500 and
HD\,179406; the latter is located closer to the Galactic Plane.}
\end{figure}

The DIBs weakly correlate with the Na\,{\sc i} absorption, but with a lot of
scatter; this indicates that in the typical warm neutral/weakly-ionized medium
both are seen and they thus each correlate with overall gas column density. In
figure 6 the behaviour of the 5780 and 5797 \AA\ DIBs is compared to that of
the Na\,{\sc i} D$_2$ line. There seems to be a threshold for the 5797 \AA\
DIB to be seen only above $EW($Na\,{\sc i} D$_2)>0.1$ \AA; instead, the 5780
\AA\ DIB can sometimes be seen already at $EW($Na\,{\sc i} D$_2)\sim0.05$ \AA.
While this is in keeping with the 5780 \AA\ DIB being stronger than the 5797
\AA\ DIB, the data in figure 6 clearly show the emergence of the 5797 \AA\ DIB
above the threshold whereas the 5780 \AA\ DIB appears to reach zero strength
at zero Na\,{\sc i} D$_2$ absorption. This observation may be understood if
the 5797 \AA\ DIB needs a minimum gas column density for the environment to be
sufficiently cool and/or shielded from UV radiation, whilst the 5780 \AA\ DIB
(and Na\,{\sc i}) are seen also in more exposed diffuse ISM.

%
\begin{figure}
\centerline{\vbox{
\epsfig{figure=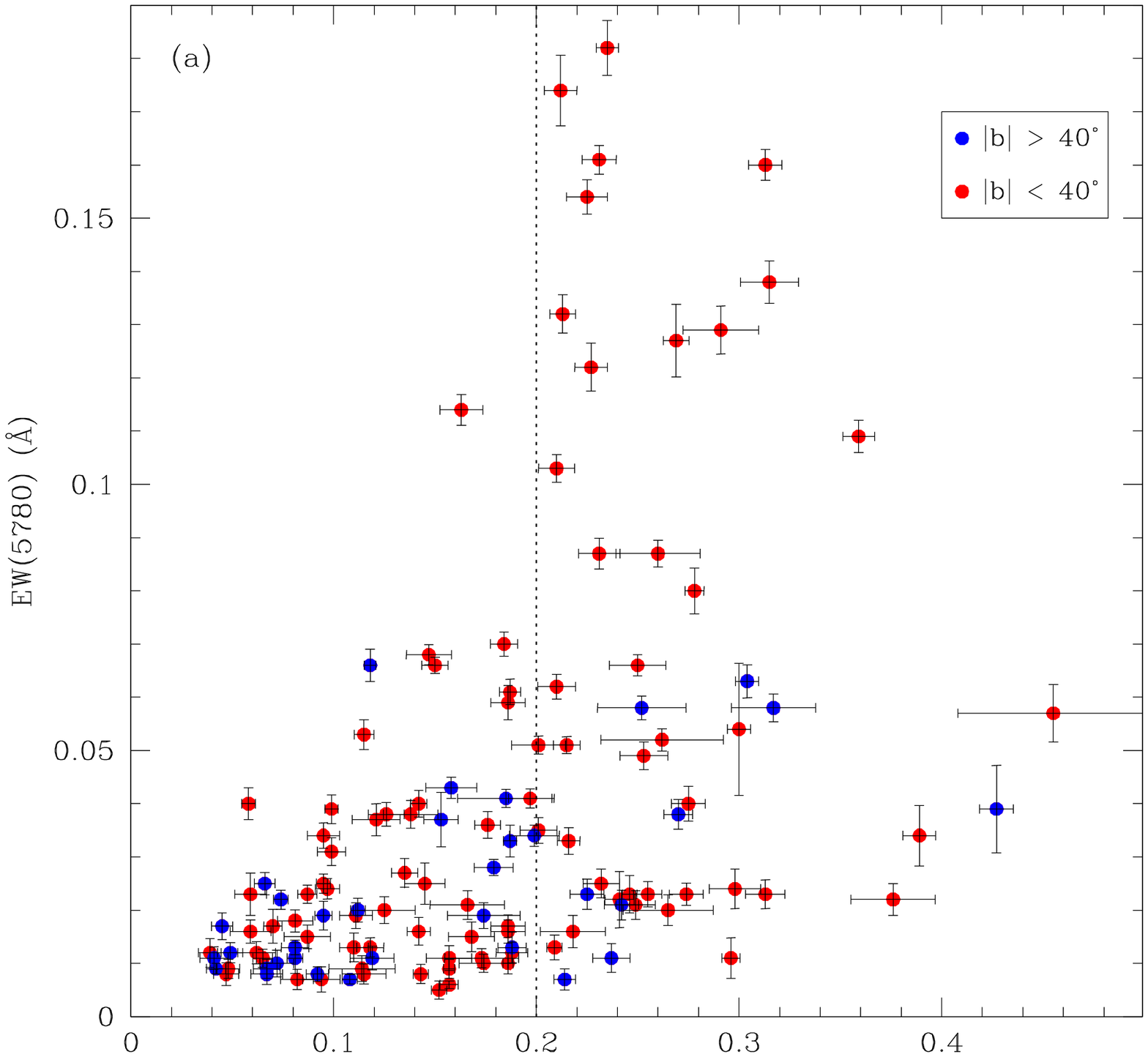,width=84mm}
\epsfig{figure=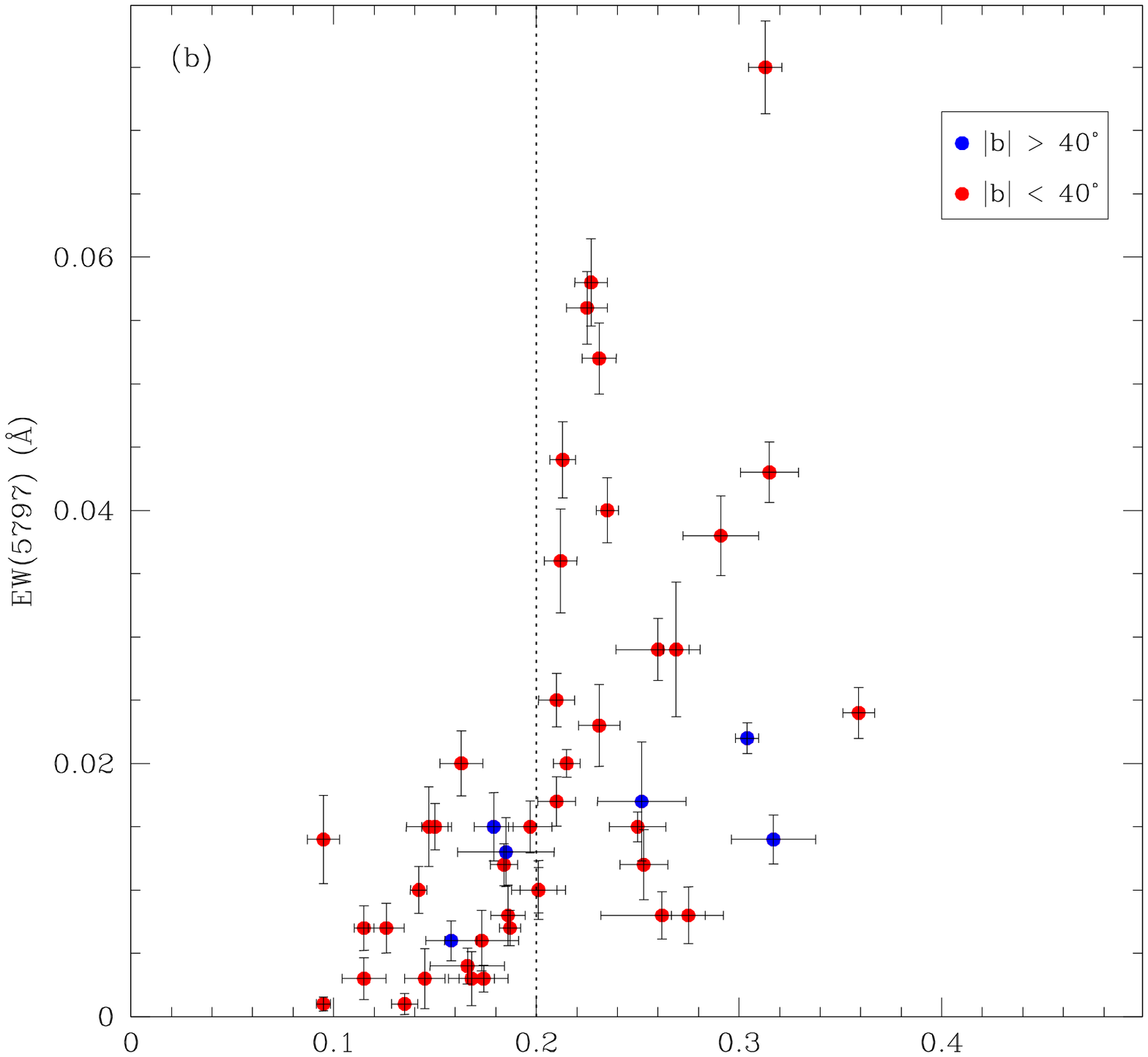,width=84mm}
\epsfig{figure=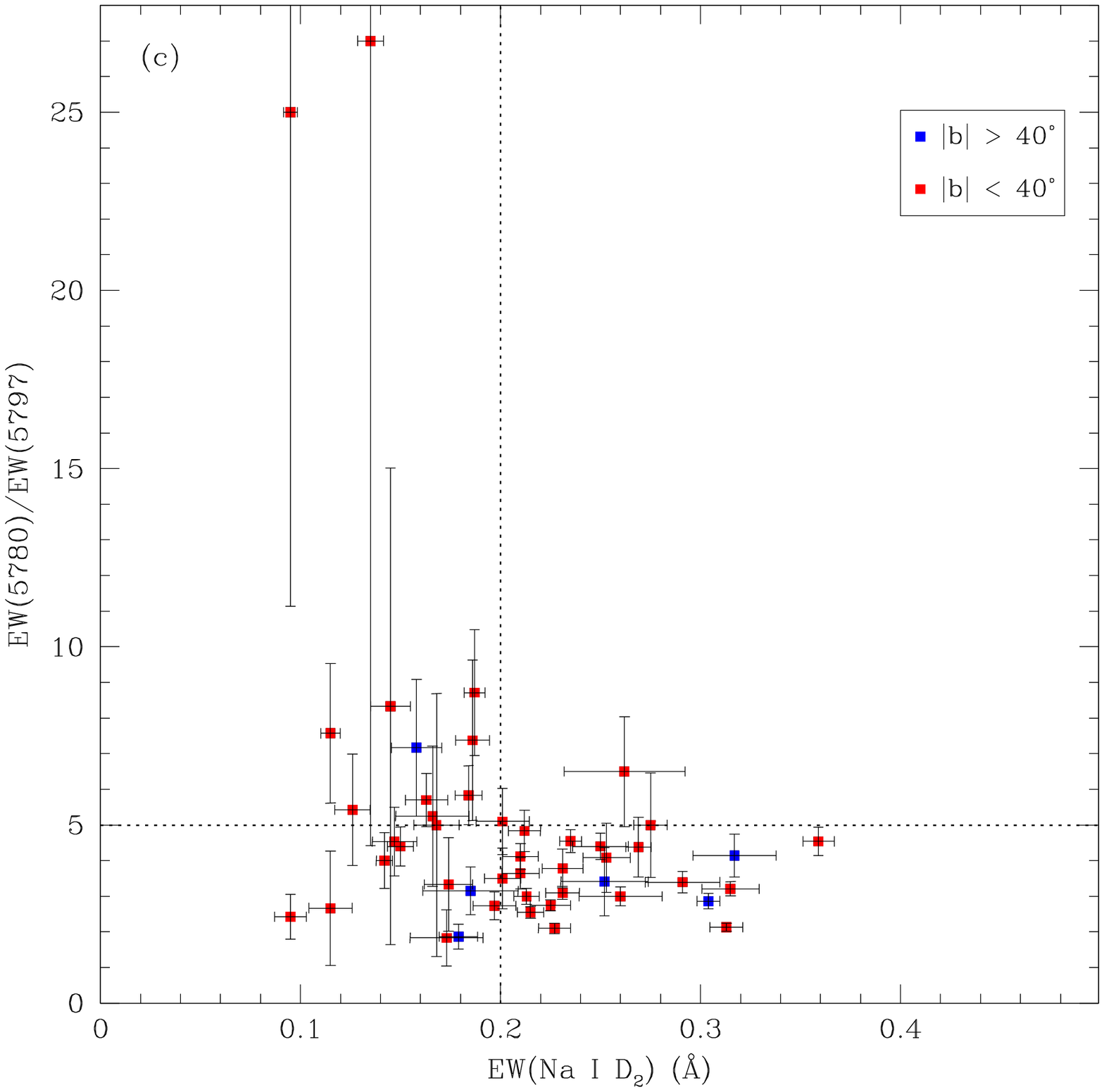,width=84mm}
}}
\caption[]{Correlations between the 5780 and 5797 \AA\ DIBs, and Na\,{\sc i}
D$_2$. Distinction is made according to Galactic latitude.}
\end{figure}

Curiously, at moderate $EW($Na\,{\sc i} D$_2)\sim0.2$--0.3 \AA\ the DIBs --
especially the 5780 \AA\ one -- are particularly strong. Certainly the 5780
\AA\ DIB is not anymore as strong once $EW($Na\,{\sc i} D$_2)\sim0.4$ \AA.
This could correspond to a greater ubiquity of material within which the
carriers of the DIBs thrive, at moderate gas densities, with the 5780 \AA\ DIB
disappearing at high(er) densities. The strong DIB absorption is, however,
only seen close to the Galactic plane, within about $|b|<40^\circ$ (cf.\ Figs.\
6a,b).

The $EW(5780)/EW(5797)$ ratio shows an intriguing behaviour with respect to
the Na\,{\sc i} absorption (Fig.\ 6c): values $EW(5780)/EW(5797)<5$ are seen
across the entire range of Na\,{\sc i} strength, but values
$EW(5780)/EW(5797)>5$ are exclusively -- and commonly -- seen only for
$EW($Na\,{\sc i} D$_2)<0.2$ \AA. Smith et al.\ (2013) find a similar extreme
$\sigma$-cloud behaviour towards $\kappa$\,Velorum (not part of our sample) at
a distance of $\sim165$ pc ($EW(5780)/EW(5797)>22$ at $EW($Na\,{\sc i}
D$_2)=0.12$ \AA). This can be understood if the Local Bubble is traced by
$EW($Na\,{\sc i} D$_2)<0.2$ \AA, where harsh UV irradiation is prevalent.
Conversely, we may assume $EW($Na\,{\sc i} D$_2)=0.2$ \AA\ as a proxy for the
boundary between the Local Bubble and surrounding ISM. The sight-lines with
$EW($Na\,{\sc i} D$_2)<0.2$ \AA\ but $EW(5780)/EW(5797)<5$ are interesting as
these may identify compact cloudlets with modest column density but high
volume density.

The huge scatter in the above correlation diagrams means that Na\,{\sc i} is
not a very good indicator of DIB carrier abundance (Merrill \& Wilson 1938;
Herbig 1995). The reason for this is partly due to the ionization potential of
sodium; neutral sodium will trace weakly ionized gas as well as neutral gas.
Therefore, it is not a very good discriminator between the two types of
environment. DIBs, on the other hand, do discriminate between the two
environments with the 5780 \AA\ DIB favouring the weakly ionized gas and the
5797 \AA\ DIB favouring the neutral gas. The use of the Na\,{\sc i}\,D lines
is further complicated by saturation effects already kicking in around
$EW($Na\,{\sc i} D$_2)>0.05$ \AA\ and the Na\,{\sc i}\,D lines not tracing
hydrogen column density in a linear fashion (Welty \& Hobbs 2001). Hence, DIBs
are a more discerning probe of the environment and give a more detailed
picture of the diffuse ISM than Na\,{\sc i} can, justifying our aim to map the
Local Bubble in DIBs (cf.\ Baron et al.\ 2015).

\subsection{Maps}

%
\begin{figure*}
\centerline{\hbox{
\epsfig{figure=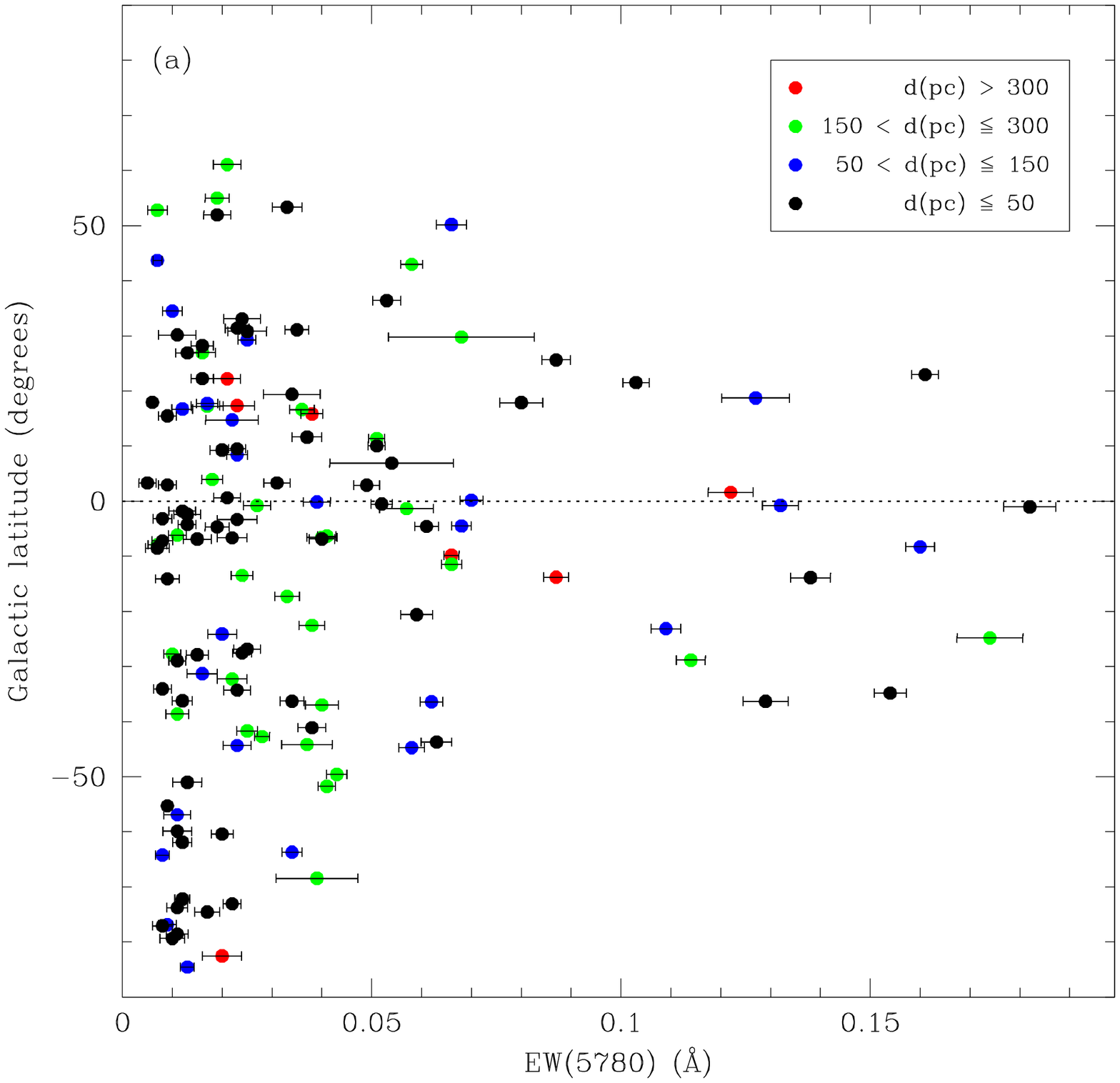,width=60mm}
\epsfig{figure=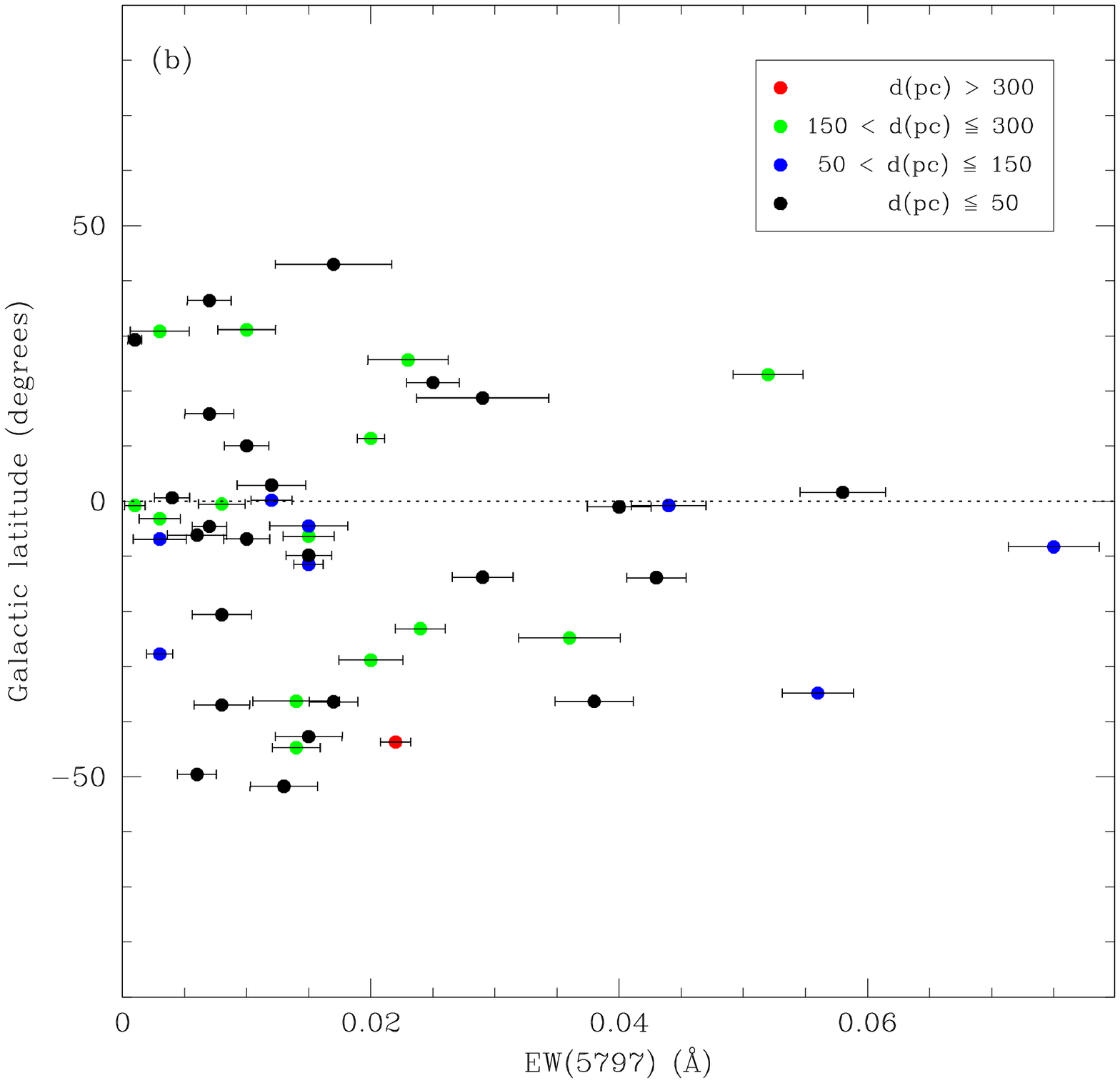,width=60mm}
\epsfig{figure=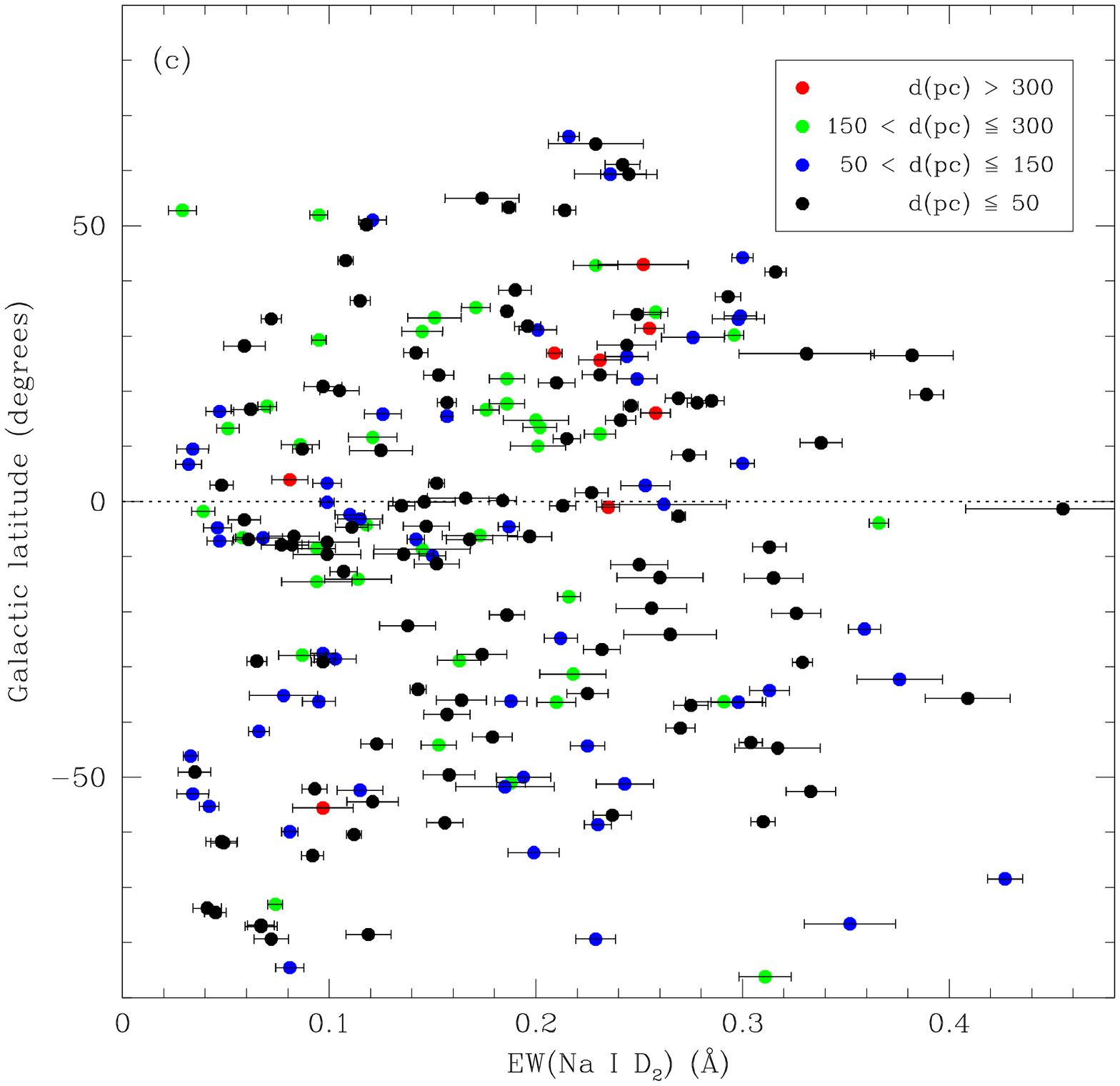,width=60mm}
}}
\caption[]{Galactic latitude distribution vs.\ equivalent width of the 5780
and 5797 \AA\ DIBs and Na\,{\sc i} D$_2$. Distinction is made according to
distance.}
\end{figure*}

%
\begin{figure*}
\centerline{\hbox{
\epsfig{figure=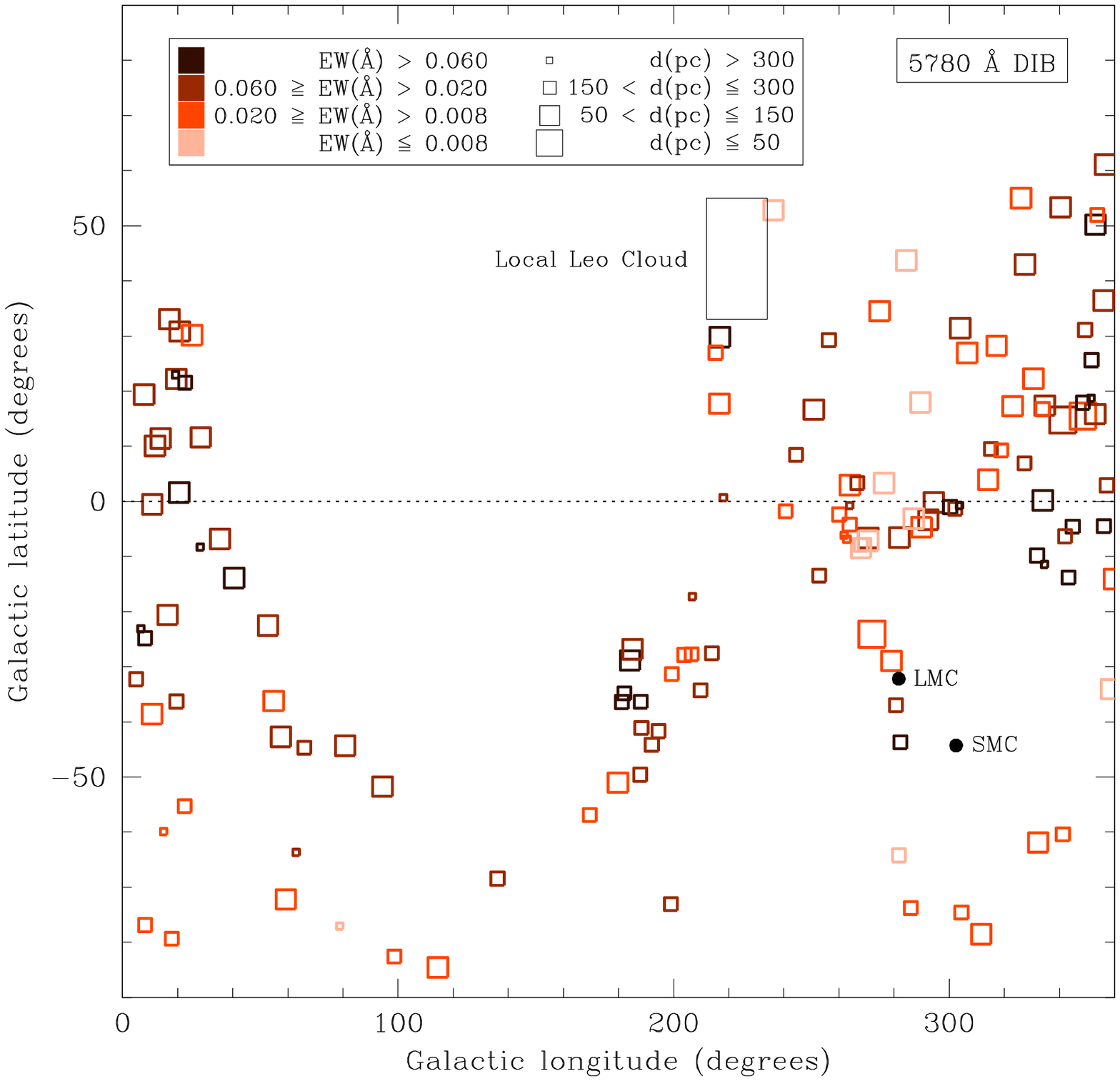,width=60mm}
\epsfig{figure=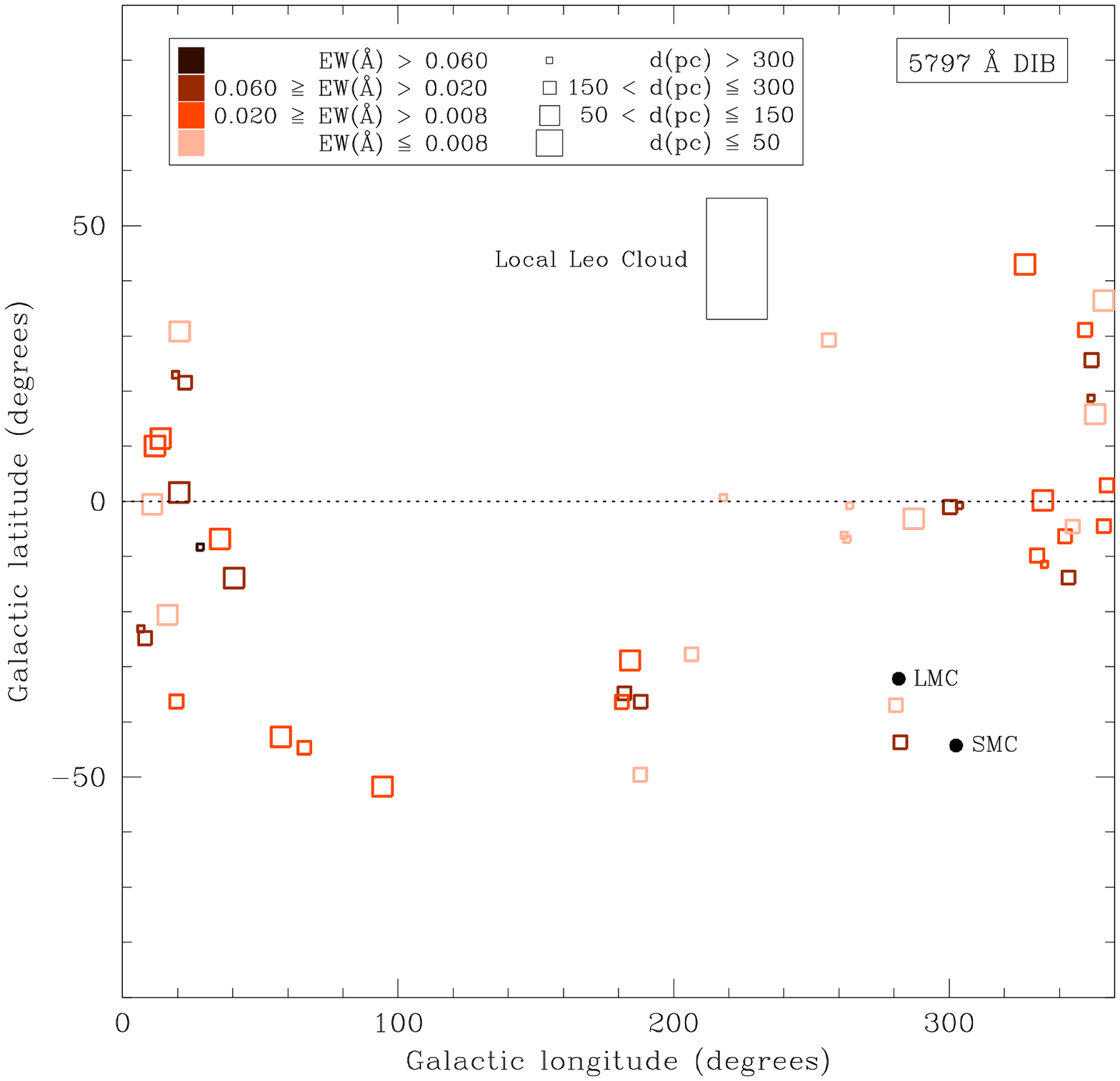,width=60mm}
\epsfig{figure=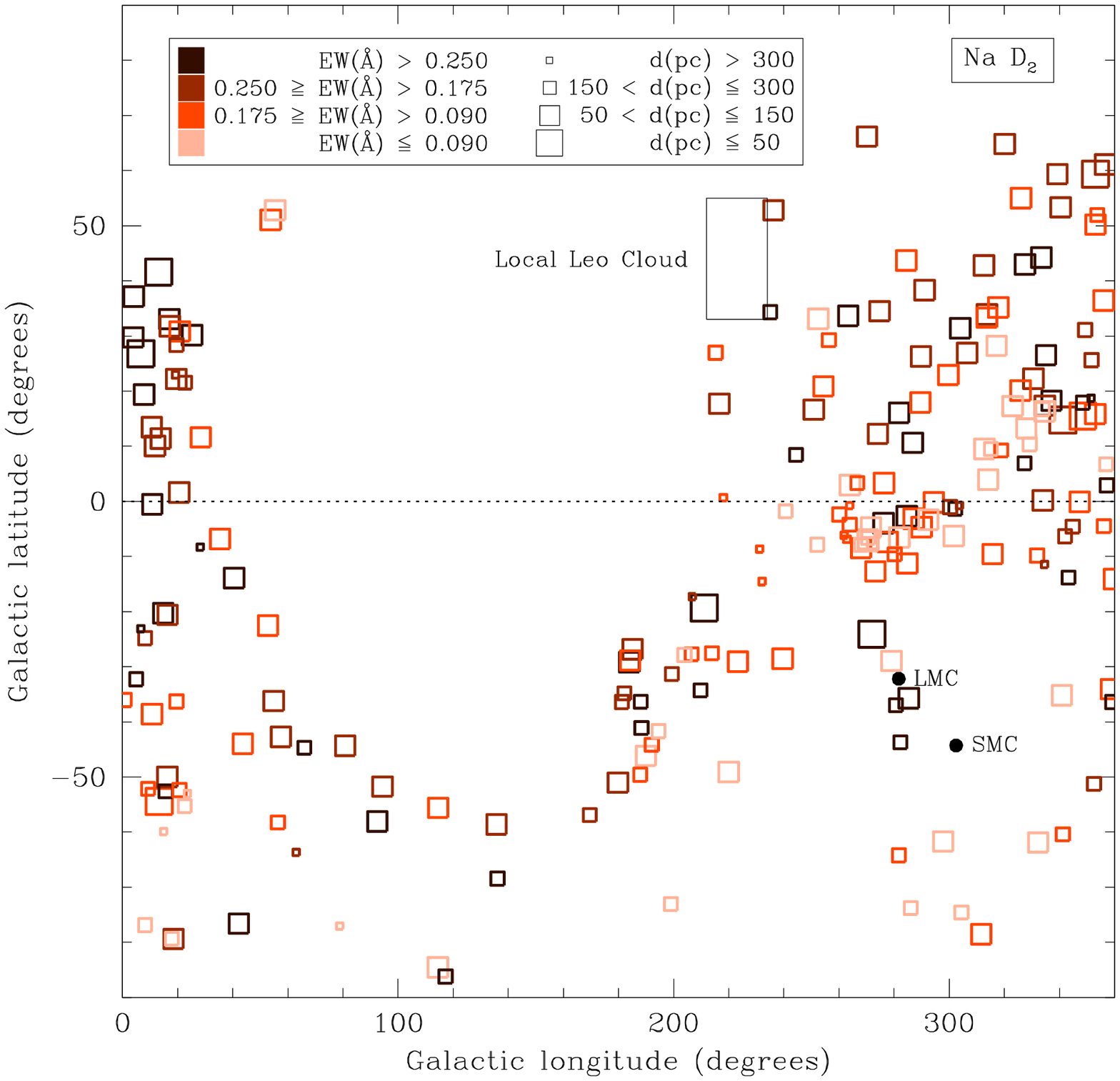,width=60mm}
}}
\caption[]{Distribution on the sky in Galactic coordinates, of the 5780 and
5797 \AA\ DIBs and Na\,{\sc i} D$_2$. The $EW$ values are colour-coded and the
distance to the Sun is indicated by the size of the symbol (smaller if
farther). The directions towards the Small and Large Magellanic Clouds and
Local Leo Cloud are also indicated.}
\end{figure*}

The Galactic latitude distribution of equivalent widths is shown in figure 7
for the 5780 and 5797 \AA\ DIBs and Na\,{\sc i} D$_2$. The latter shows no
relation whatsoever, but the DIBs clearly are stronger towards the Galactic
Plane, within about $|b|<40^\circ$. The lack of dependence of Na\,{\sc i} may
be due in part to saturation though stronger Na\,{\sc i} is seen towards more
distant objects than in our survey (e.g., van Loon et al.\ 2009). Neither of
these three tracers shows clear dependence on distance (colour-coded in the
graphs). This implies that within a few hundred pc from the Sun, small-scale
structure dominates the variations between different sight-lines. As the DIBs
do ``detect'' the Galactic Plane they seem to be more useful tracers to map
the Local Bubble than Na\,{\sc i}.

To add a dimension, in figure 8 are plotted the sight-lines in Galactic
coordinates, i.e.\ projected on the sky, with $EW$ values and distances to
the Sun colour-coded and indicated by the size of the symbol, respectively.
Thus we start to gain a more three-dimensional picture of the absorption. In
these projections, there is a clear gap which corresponds to the Northern
hemisphere which is covered in Papers II \& III. Also indicated are the
directions towards the Magellanic Clouds, which have been studied in the past
(e.g., Cox et al.\ 2006; van Loon et al.\ 2013) and is the subject of a large
DIB survey presented in Bailey et al.\ (2015).

Of more immediate relevance is the location of the Local Leo Cloud, a nearby
very cold cloud ($T\sim20$ K) between 11.3--24.3 pc from the Sun. Hence it is
well within the cavity of the Local Bubble and the closest known cold neutral
gas cloud (Peek et al.\ 2011; Meyer et al.\ 2012). Our targets skirt this
cloud but some sight-lines from the Northern hemisphere survey crossed it. In
the present data set two sight-lines just South of the Cloud have rather
strong 5780 \AA\ DIB absorption, without detection of the 5797 \AA\ DIB. This
could be tracing the ionized skin of the Cloud.

Overall, the strongest absorption is concentrated within $\sim\pm25^\circ$ of
the Galactic Plane (Fig.\ 8), but with exceptions. For instance, relatively
strong absorption in both the 5780 and 5797 \AA\ DIBs as well as Na\,{\sc i}
is seen around $l\sim190^\circ$ and $b\sim-25^\circ$ to $-40^\circ$. The wall
of the Local Bubble may be extending further away from the Galactic Plane,
possibly if it is relatively close to us. On the other hand, sight-lines
around $l\sim270^\circ$ are more transparent despite quite long columns --
especially when compared to the adjacent sight-lines around $l\sim300^\circ$;
perhaps the Local Bubble extends further in the disc in the former direction.

The maps in figures 9 and 10 show the $EW$ of the 5780 and 5797 \AA\ DIBs and
Na\,{\sc i} D$_2$, colour-coded, in two projections: onto the Galactic Plane
and in an orthogonal plane crossing the Galactic Centre also (the
``meridional'' plane). In figure 10 also non-detections are shown; in some
cases these would mask the detections hence the separate plots. The Sun is
marked at the coordinates (0,0). The size of the symbols represents the
distance from the projection plane, with circles representing targets that are
above/in front and squares representing targets below/behind the plane,
respectively.

%
\begin{figure*}
\centerline{\vbox{
\hbox{
\epsfig{figure=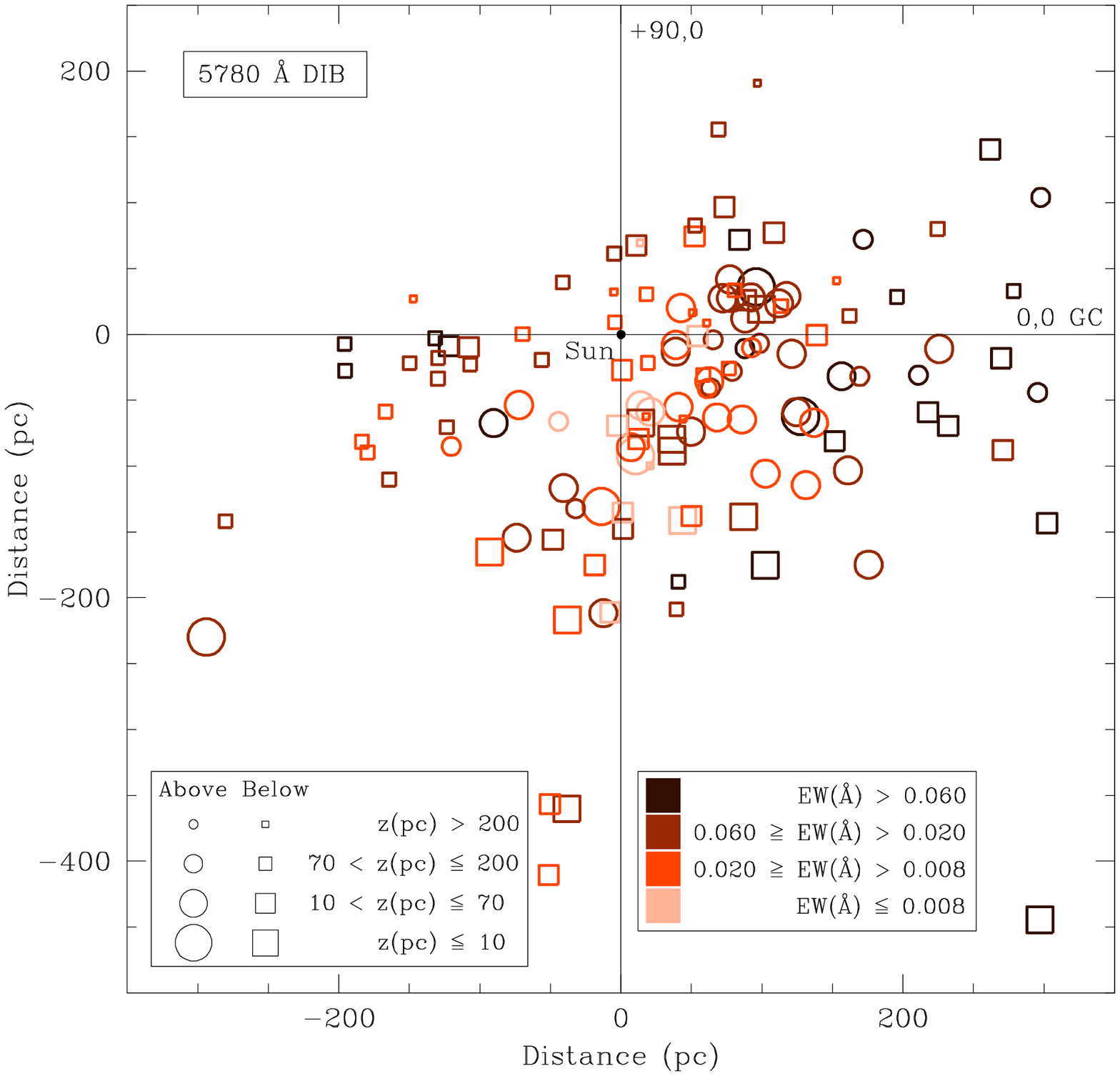,width=59mm}
\hspace{1mm}
\epsfig{figure=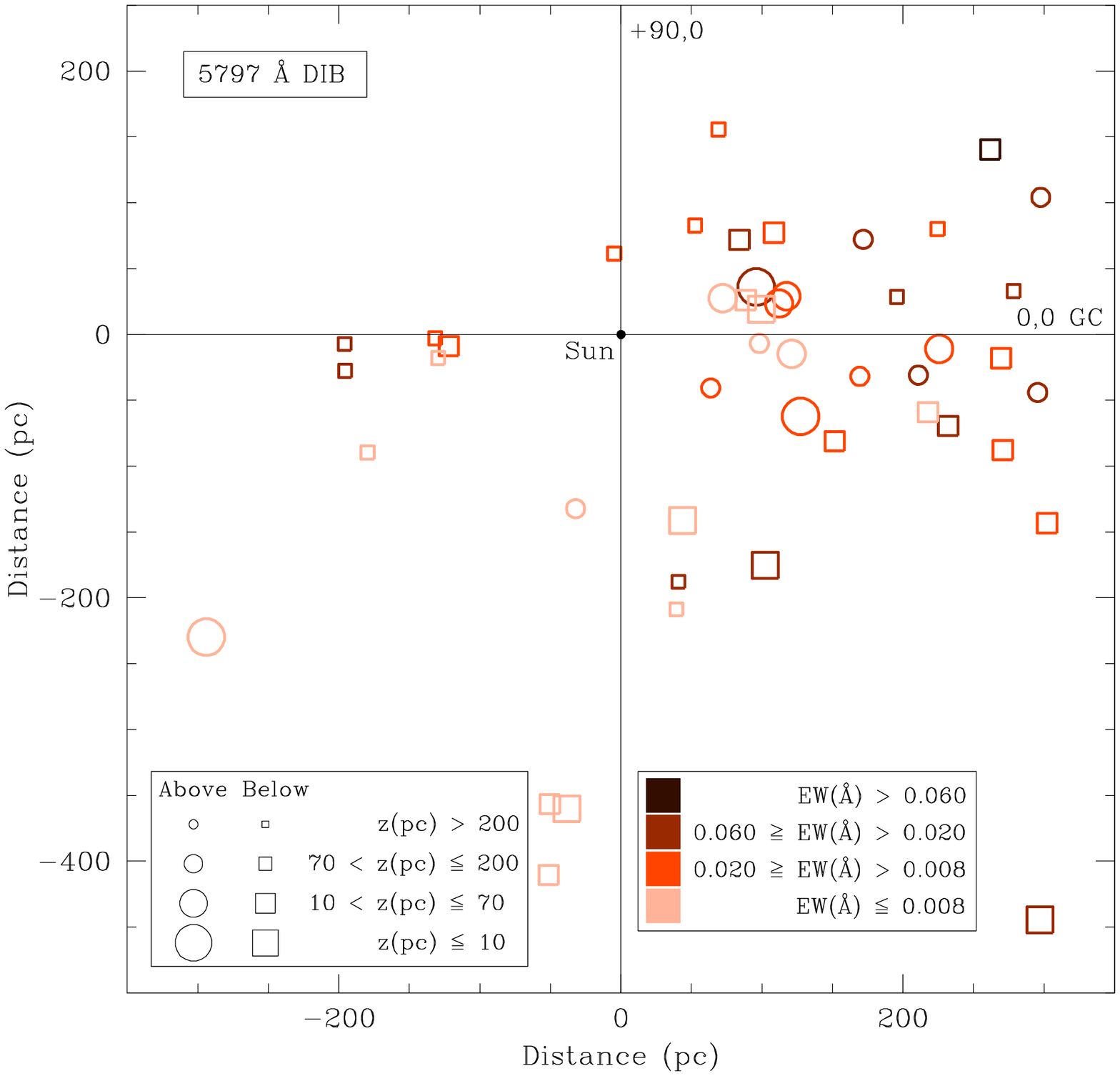,width=59mm}
\hspace{1mm}
\epsfig{figure=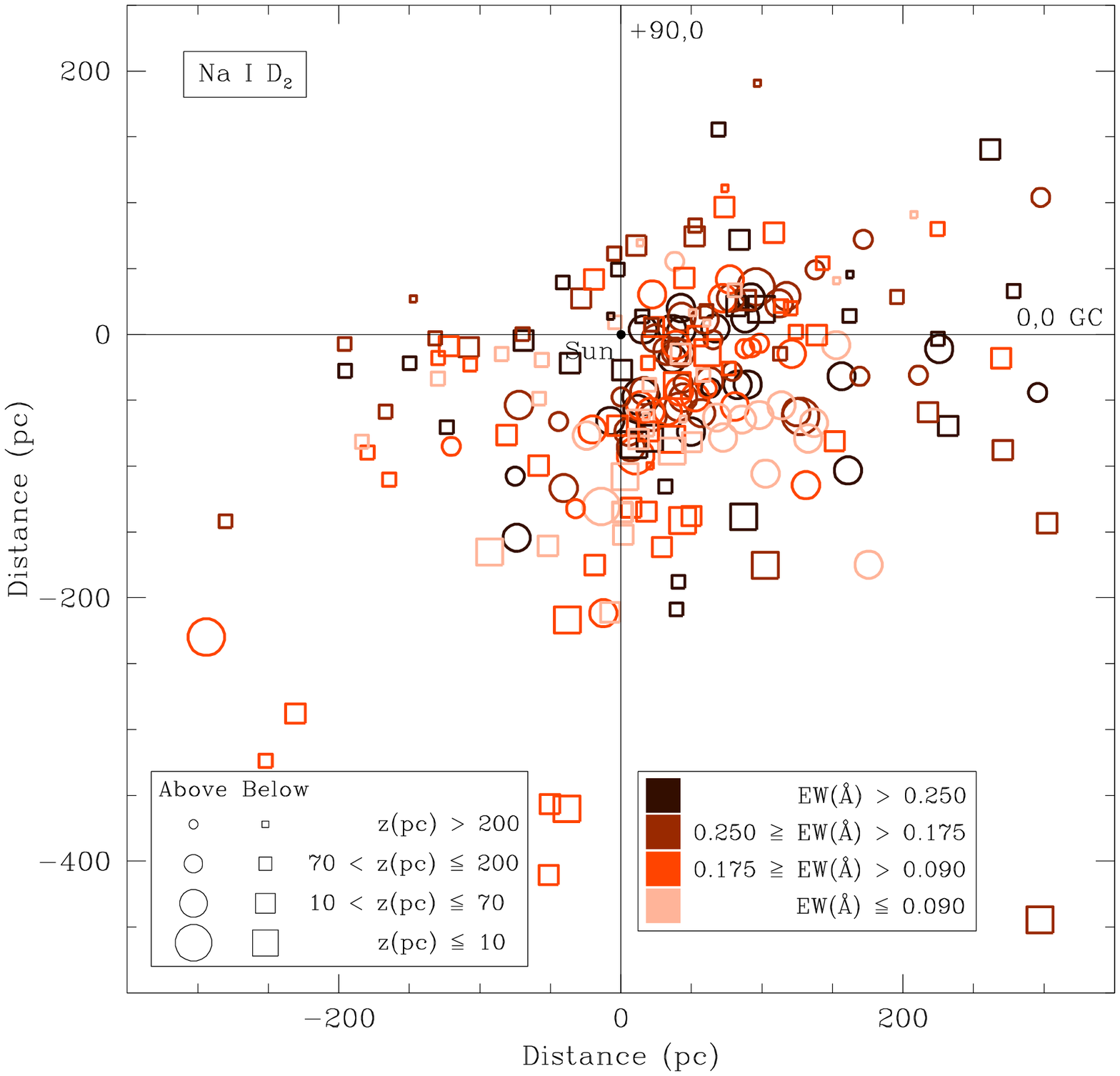,width=59mm}
}
\hbox{
\epsfig{figure=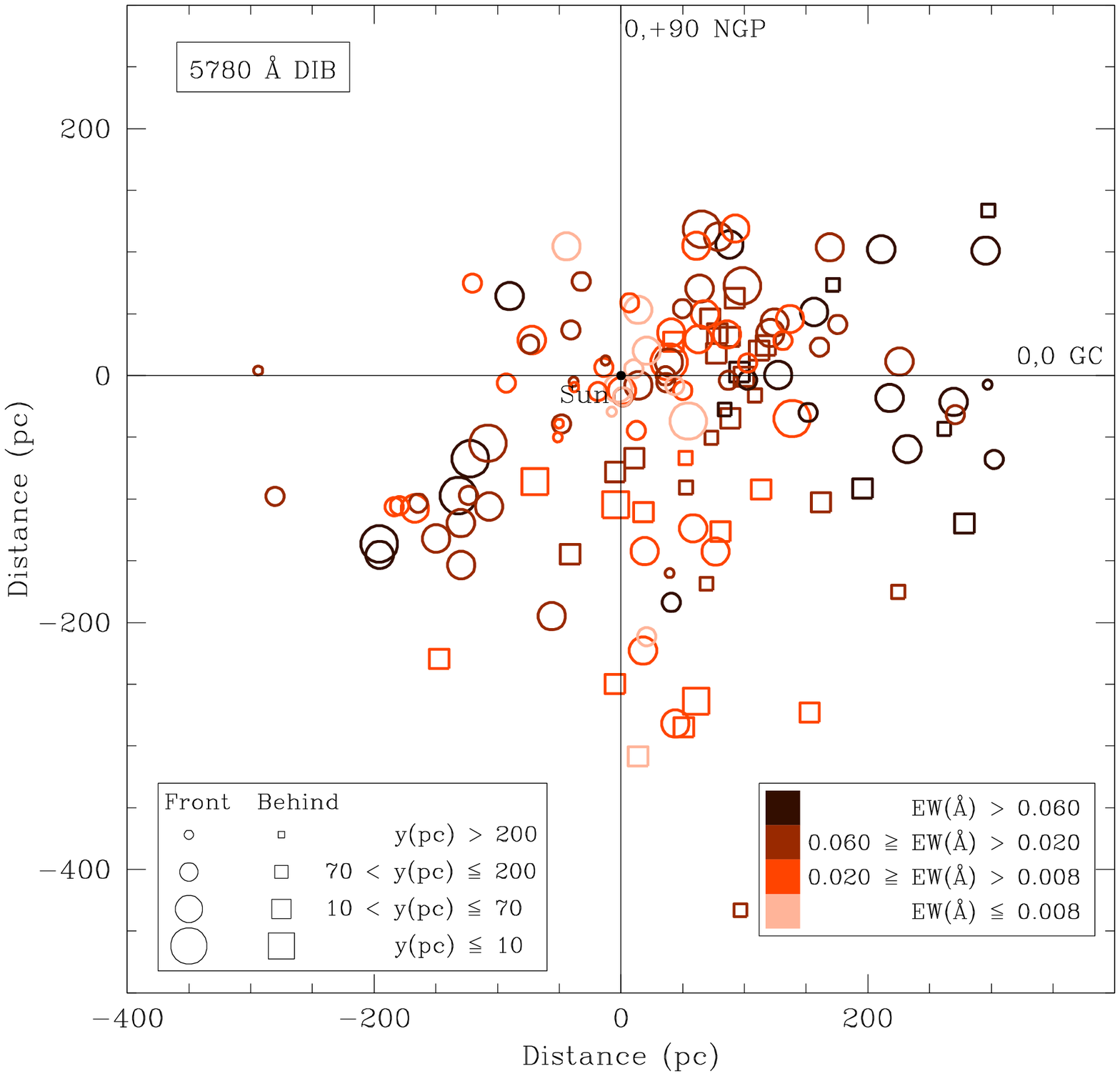,width=59mm}
\hspace{1mm}
\epsfig{figure=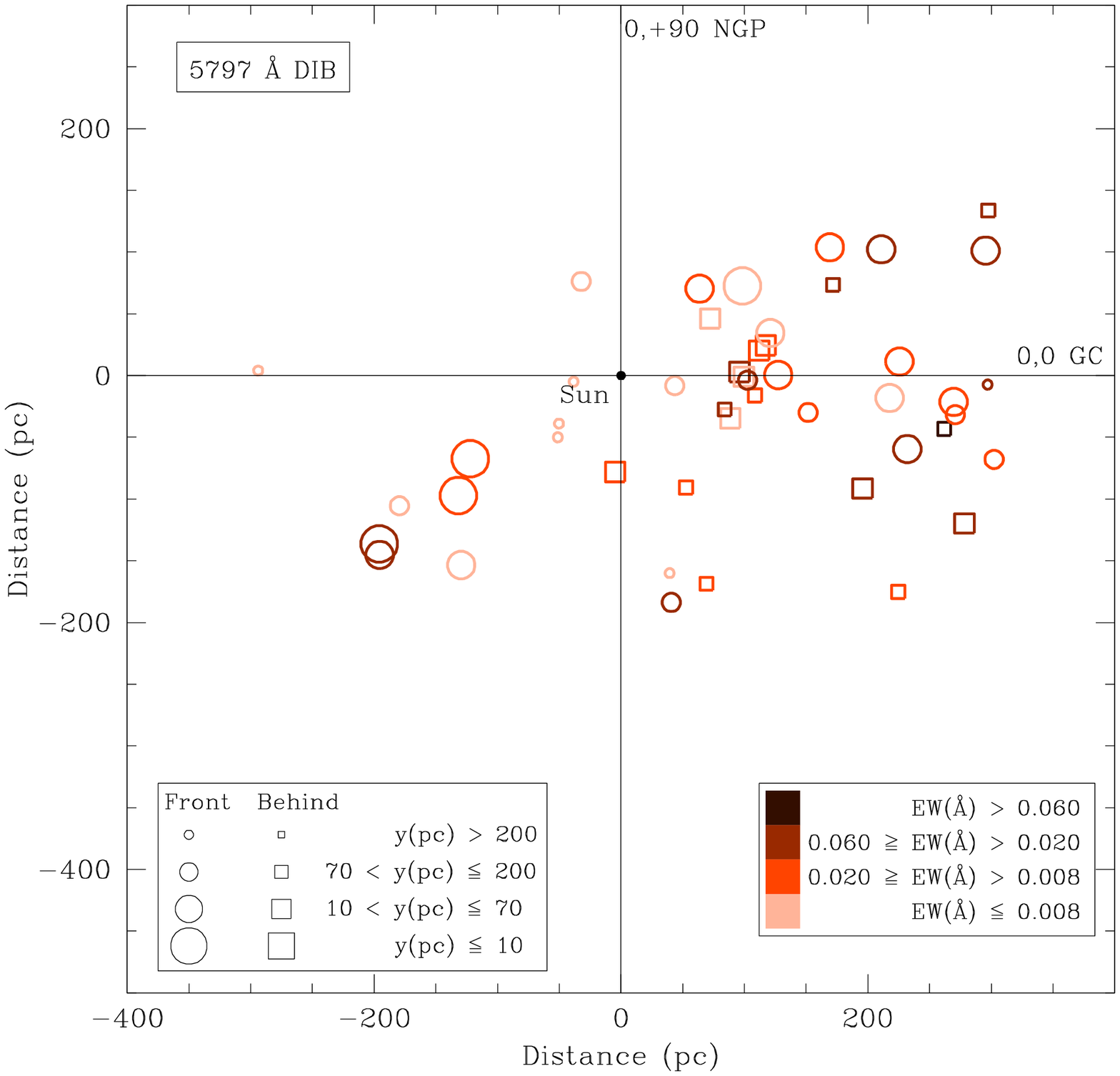,width=59mm}
\hspace{1mm}
\epsfig{figure=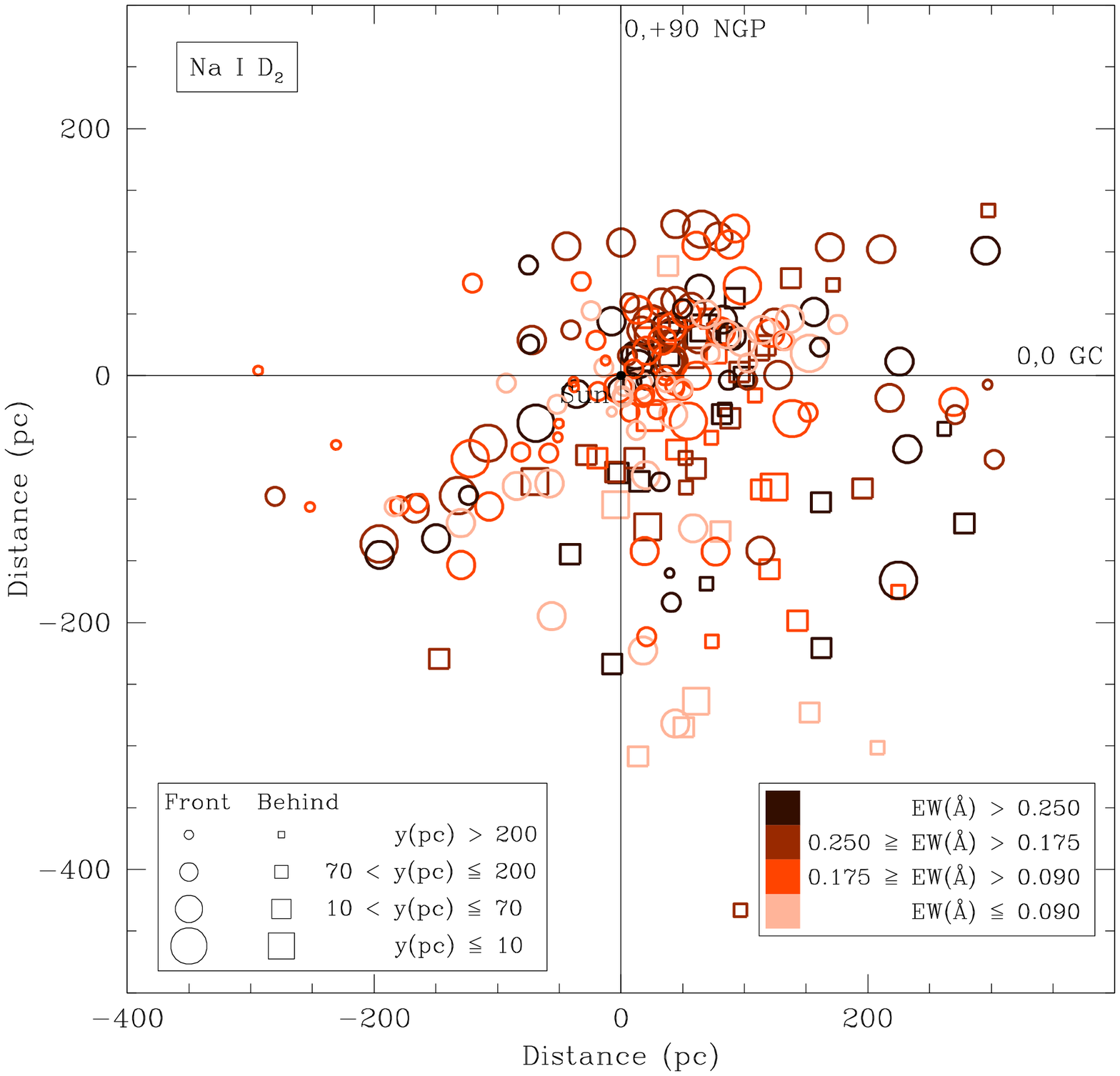,width=59mm}
}
}}
\caption[]{Galactic maps of ({\it Left to Right:}) 5780 \AA\ DIB, 5797 \AA\
DIB and Na\,{\sc i} D$_2$ absorption, in projection ({\it Top:}) onto the
Galactic Plane and ({\it Bottom:}) in the meridional plane orthogonal to the
Galactic Plane. Only detections are included in these maps.}
\end{figure*}

%
\begin{figure*}
\centerline{\vbox{
\hbox{
\epsfig{figure=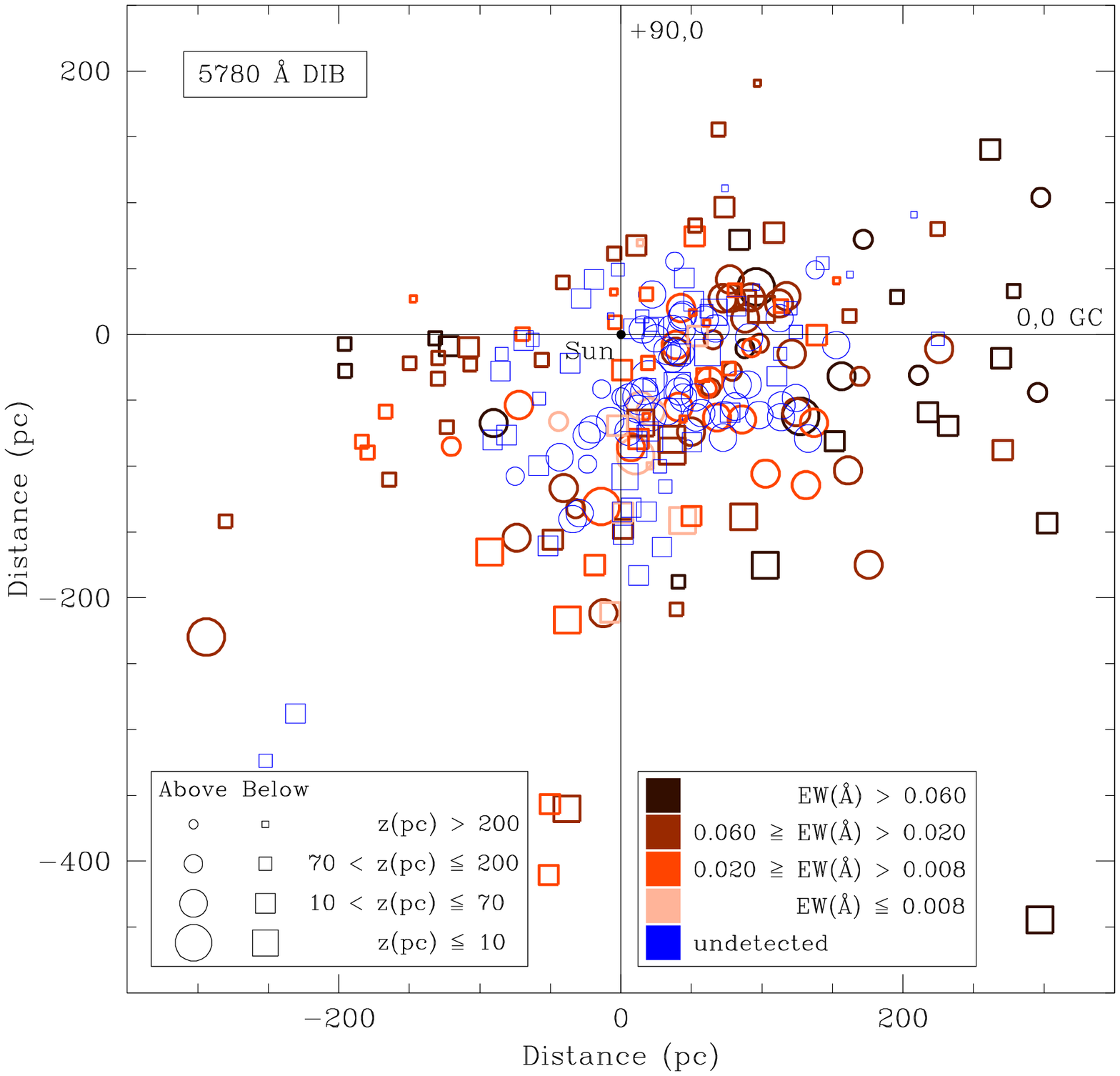,width=59mm}
\hspace{1mm}
\epsfig{figure=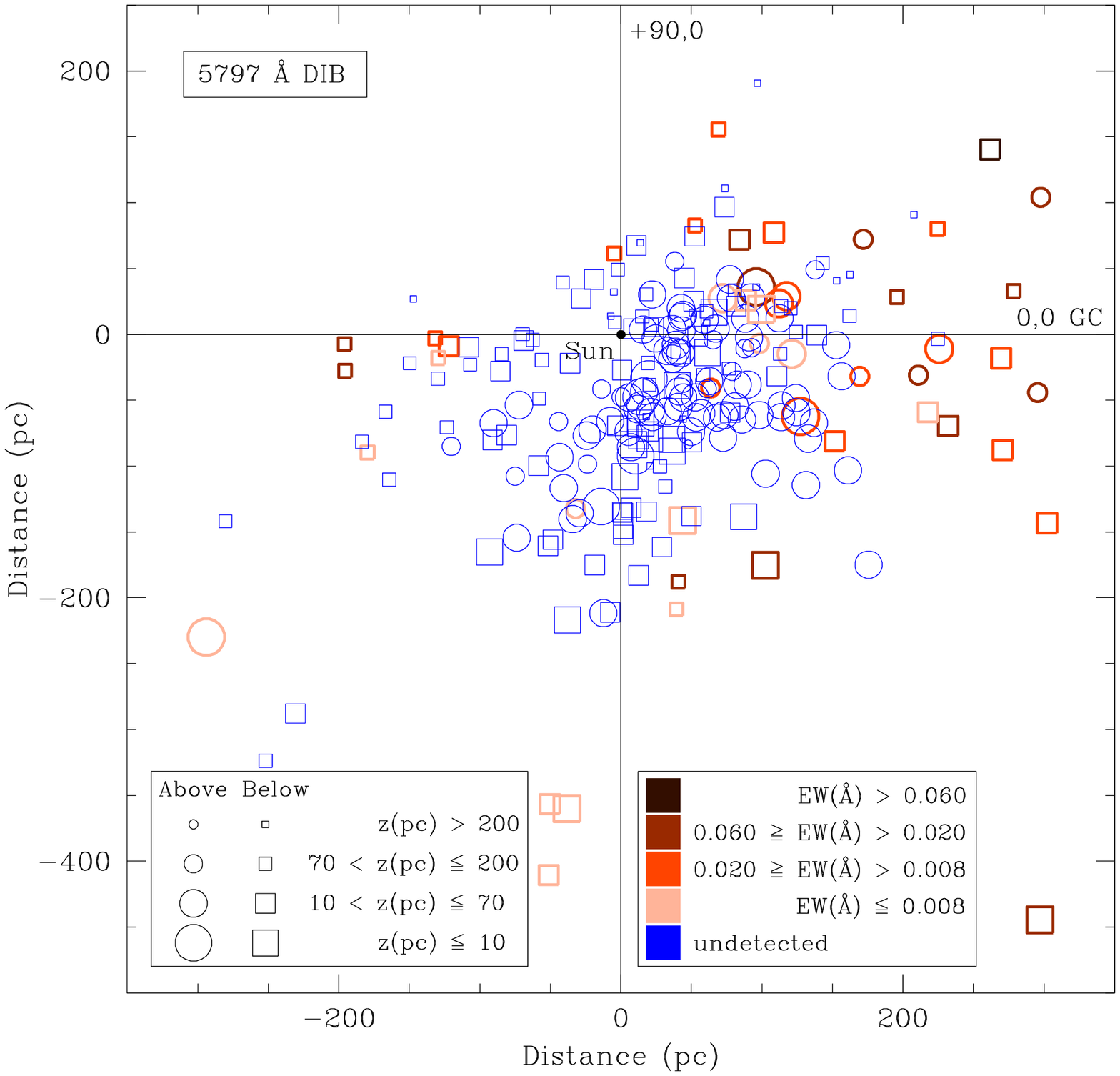,width=59mm}
\hspace{1mm}
\epsfig{figure=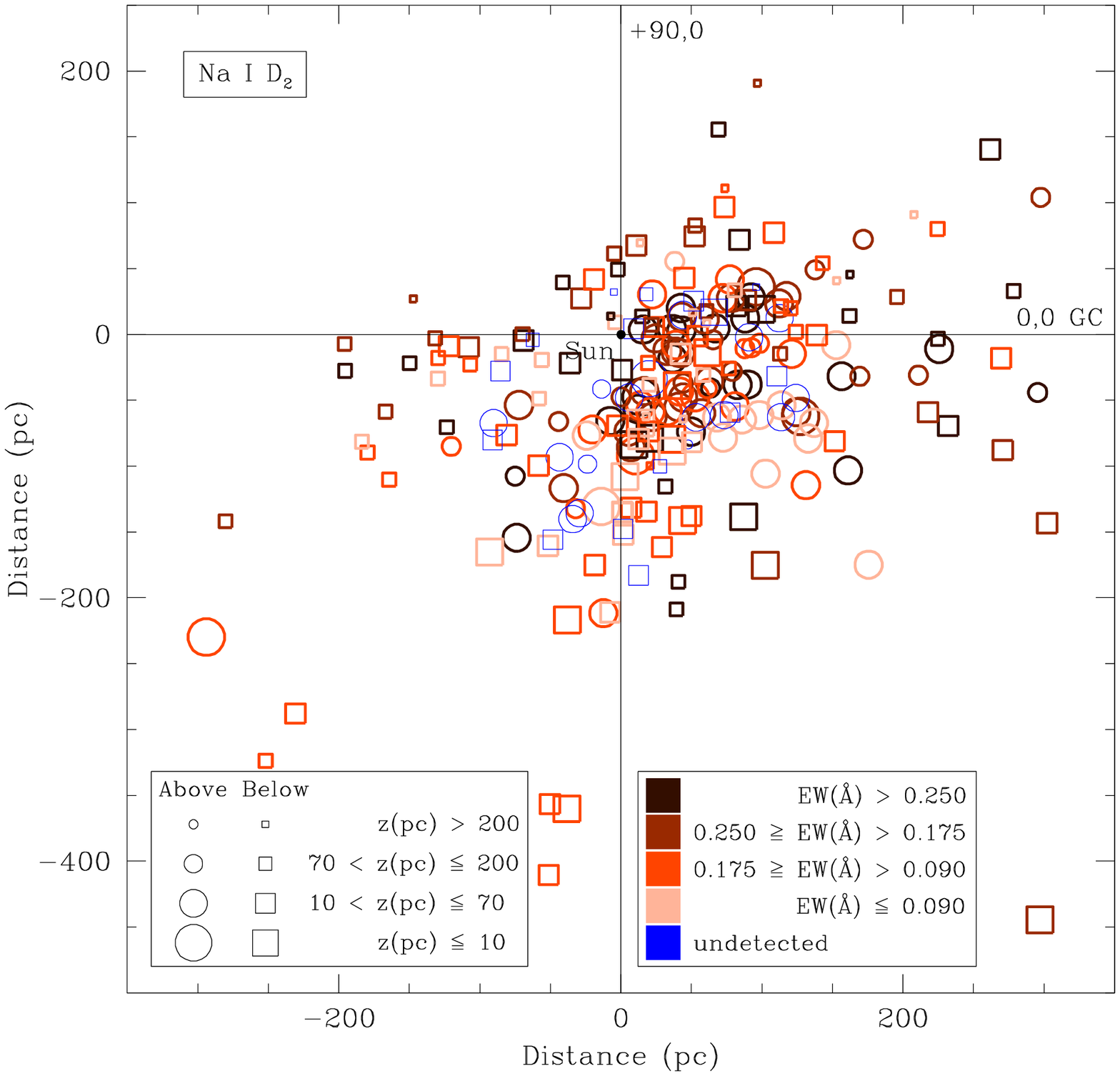,width=59mm}
}
\hbox{
\epsfig{figure=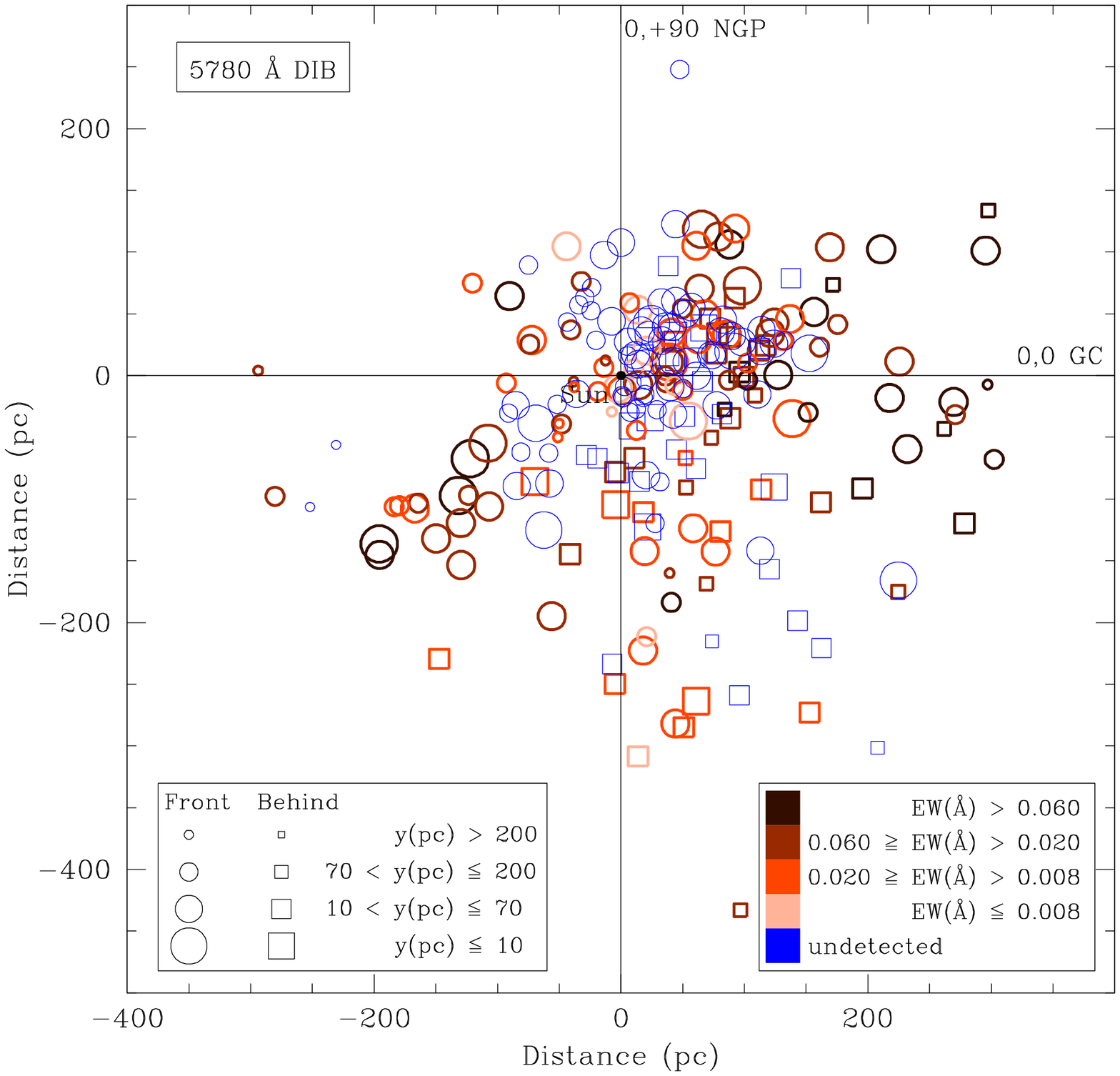,width=59mm}
\hspace{1mm}
\epsfig{figure=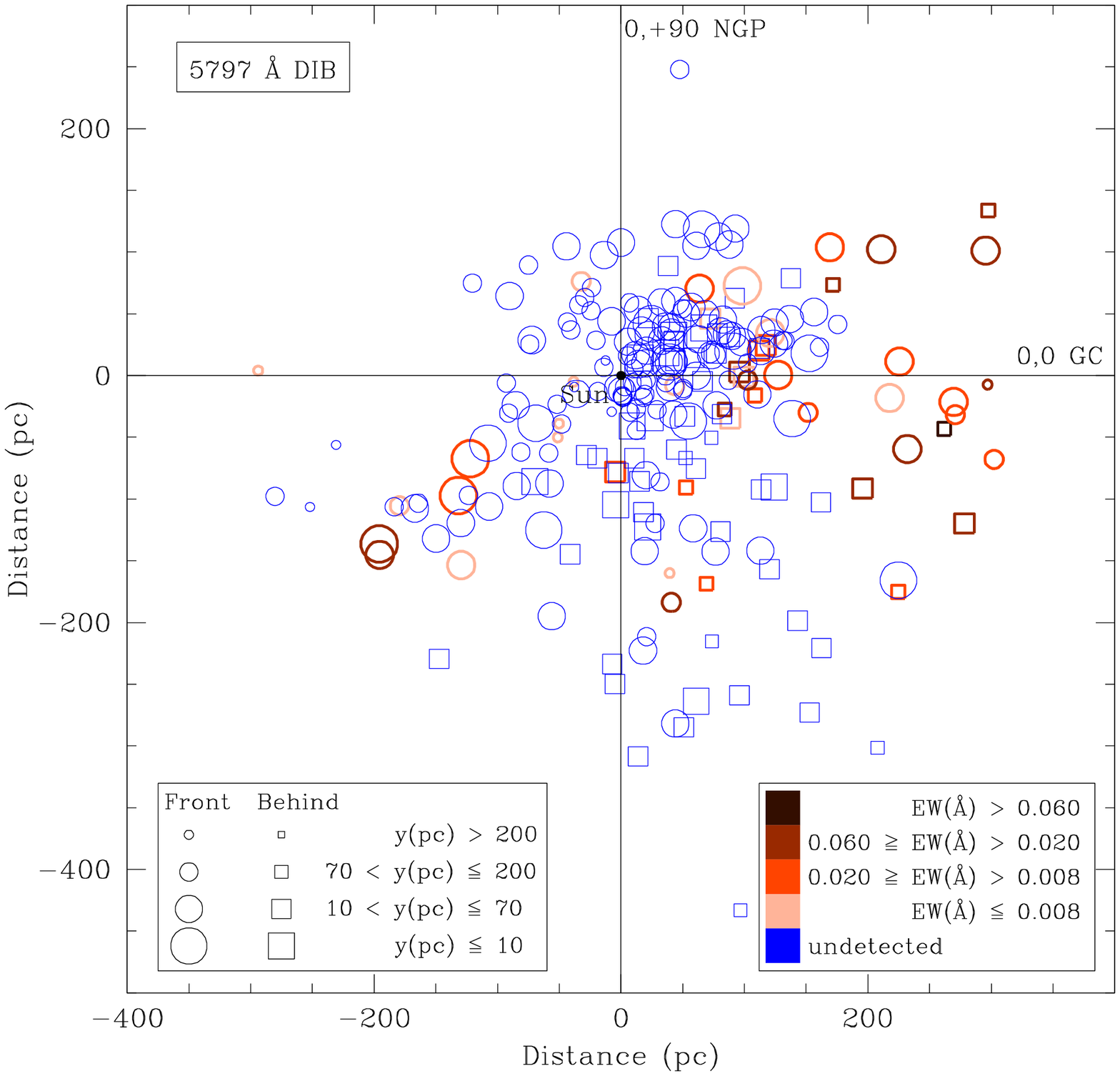,width=59mm}
\hspace{1mm}
\epsfig{figure=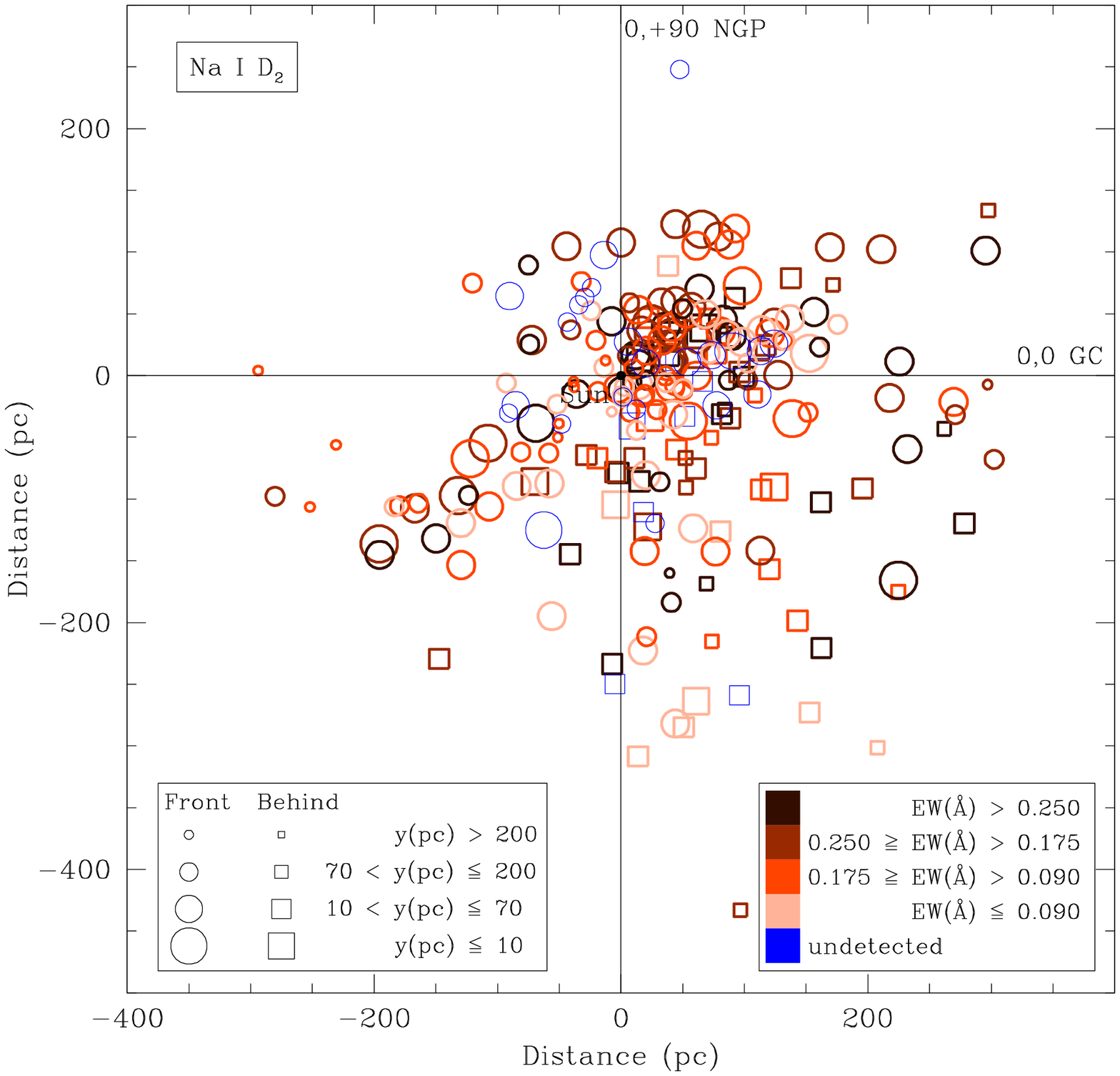,width=59mm}
}
}}
\caption[]{As figure 9, with non-detections included.}
\end{figure*}

Areas of strong 5780 and/or 5797 \AA\ DIB absorption generally match areas of
strong Na\,{\sc i} D$_2$ absorption and high Na$^0$ volume density in the
Welsh et al.\ (2010) map. There are fewer sight-lines with detected 5797 \AA\
DIB absorption, hence these maps are sparser; however, including the
non-detections, figure 10 clearly shows the extent of the Local Bubble and in
particular a ``chimney'' structure perpendicular to the Galactic Plane, with
5797 \AA\ DIB absorption confined to the Plane. The 5780 \AA\ DIB is also seen
further out along the chimney, commensurate with it being ubiquitious in the
extra-planar gas (van Loon et al.\ 2009). So the Local Bubble opens out into
the Halo in some tracers but not others. Likewise, the sight-lines between
$l\sim200^\circ$ to $270^\circ$ are transparent in the 5797 \AA\ DIB but
absorption in the 5780 \AA\ DIB is clearly seen along those same sight-lines.
This implies translucent density enhancements within the Local Bubble, or a
fuzzier, extended transition rather than a sharp boundary between the Local
Bubble and surrounding ISM. It is thus imperative not to ignore non-detections
in mapping the ISM structure, as they clearly trace cavities but also can
indicate porosity or clumpiness.

%
\begin{figure*}
\centerline{\vbox{
\hbox{
\epsfig{figure=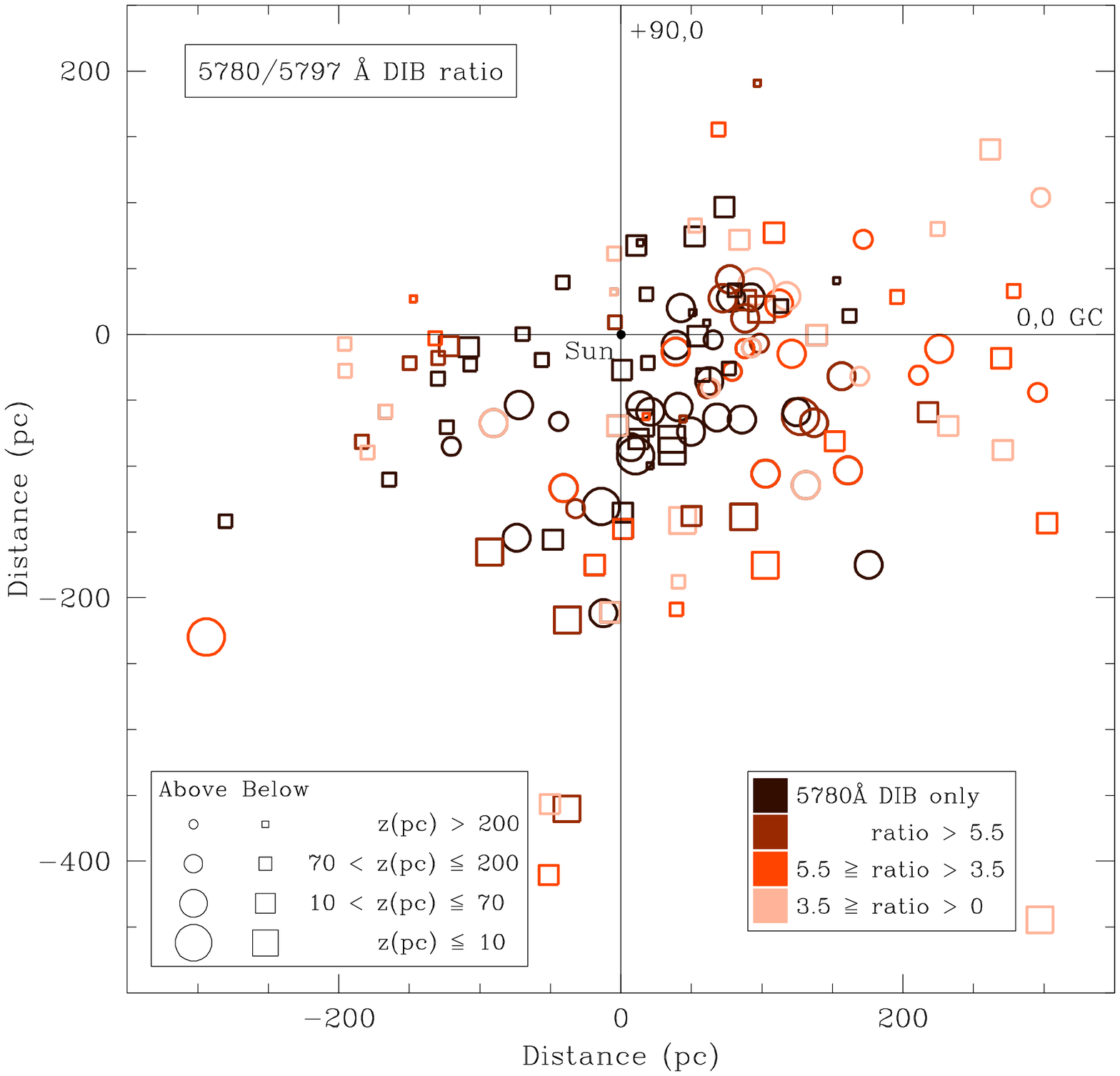,width=90mm}
\epsfig{figure=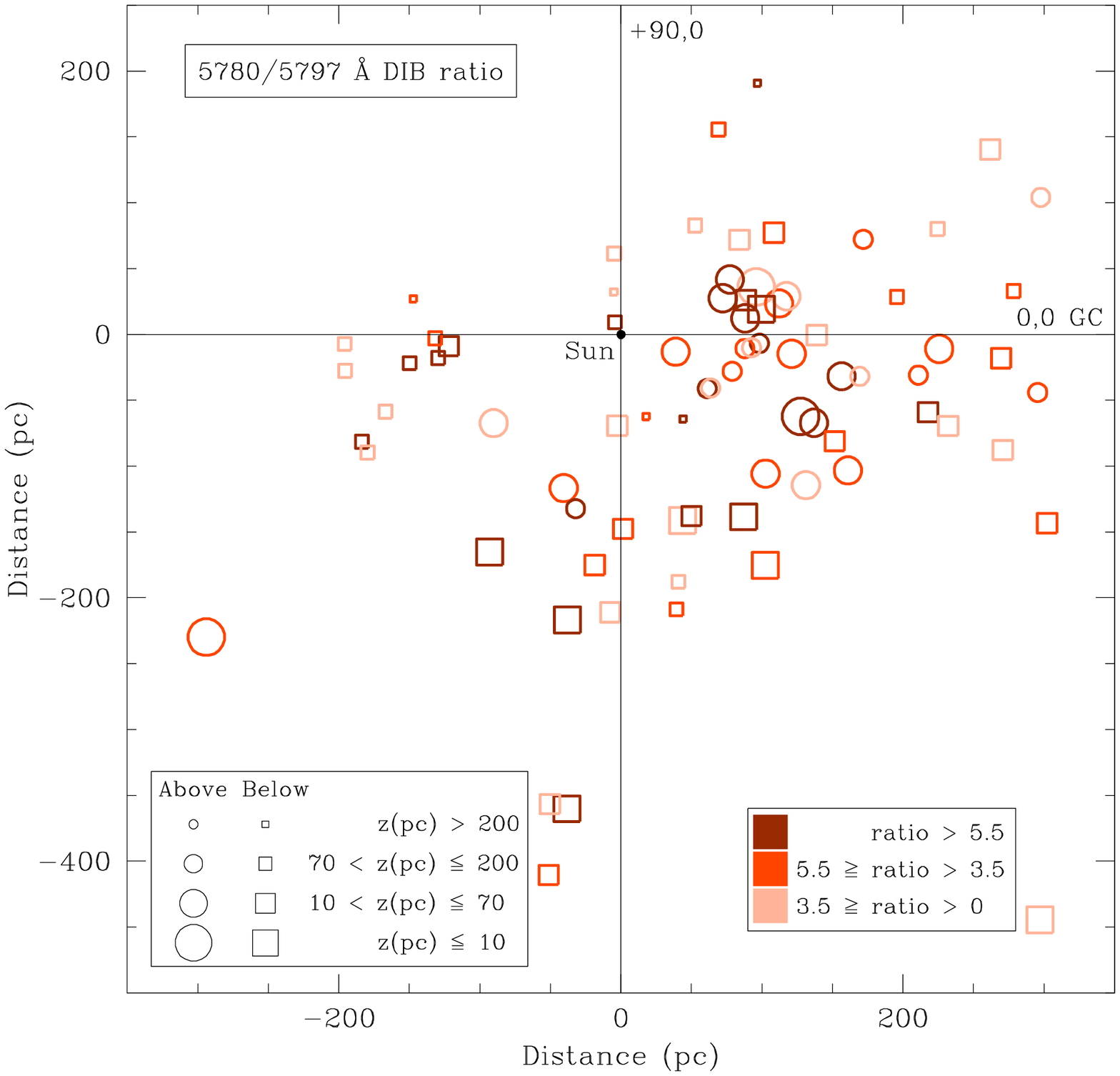,width=90mm}
}
\hbox{
\epsfig{figure=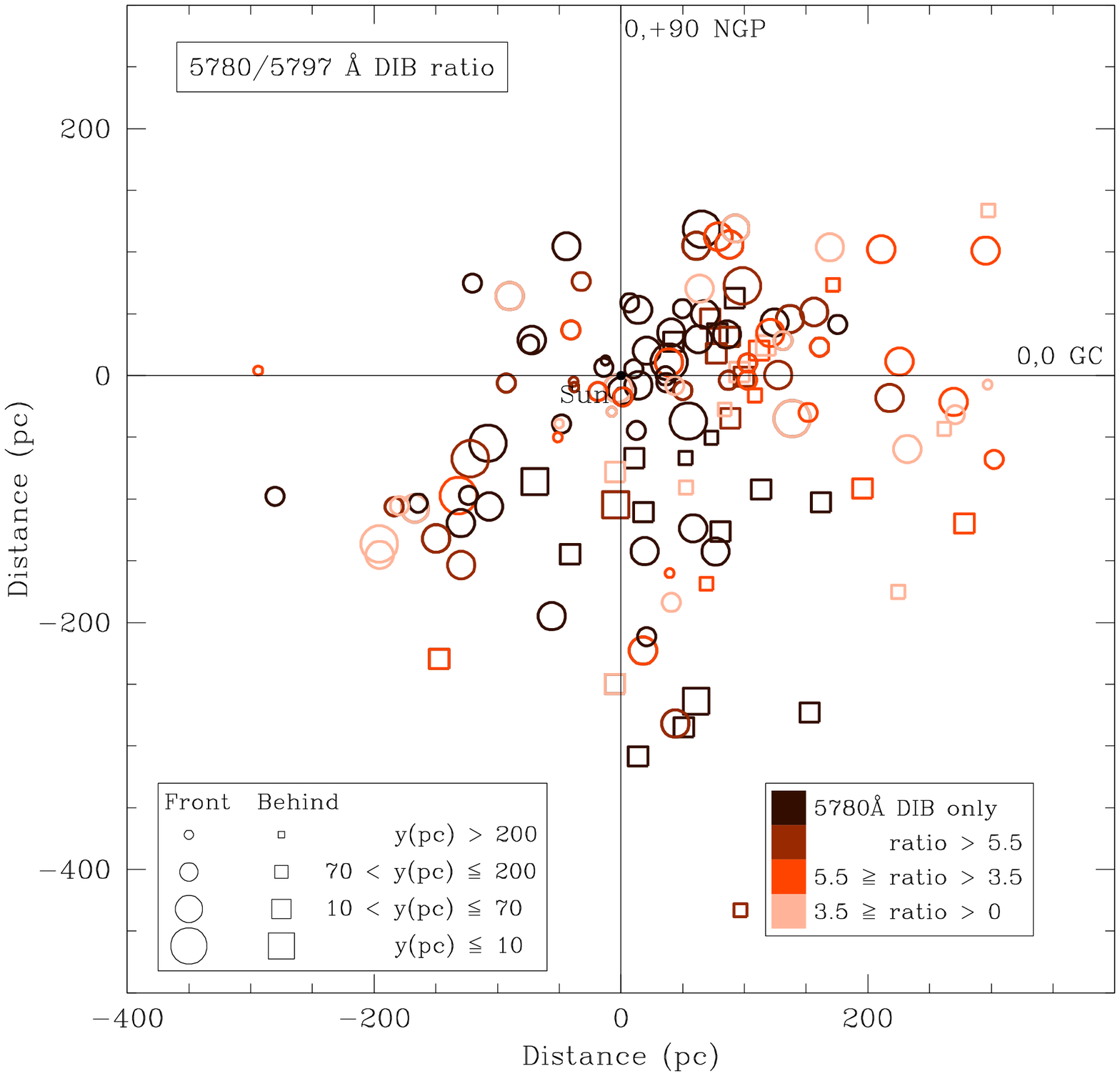,width=90mm}
\epsfig{figure=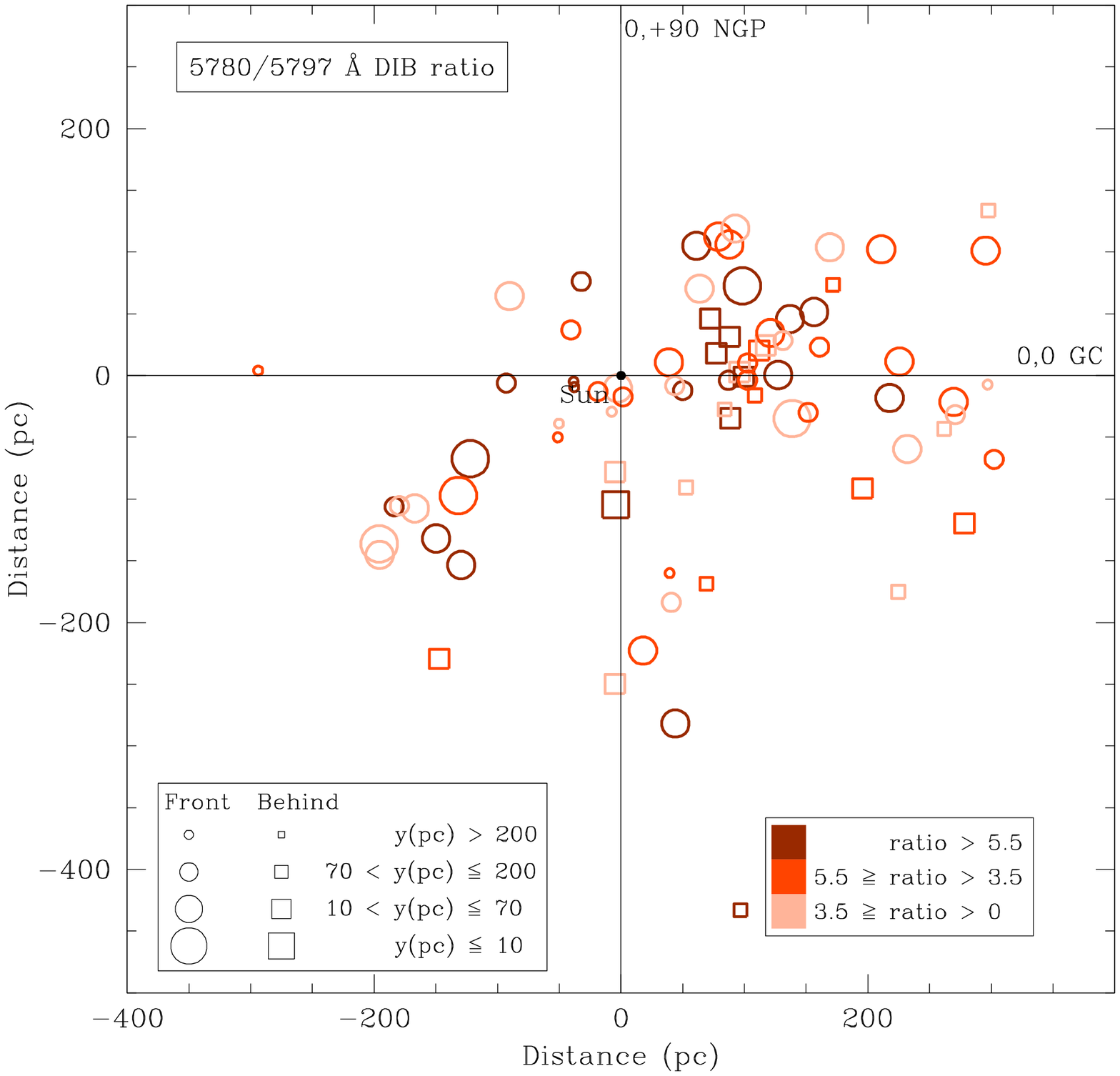,width=90mm}
}
}}
\caption[]{As figure 9, but for the 5780 to 5797 \AA\ DIB ratio. The panels on
the right only show ratios derived from detections in both DIBs.}
\end{figure*}

This is emphasized in maps of the ratio of 5780 to 5797 \AA\ DIB absorption
(Fig.\ 11), where the paler colours trace the neutral gas in the wall of the
Local Bubble and the darker colours trace the chimney. This is especially
clear in the meridional plane projection, but even in the Galactic Plane
projection the cavity of the Local Bubble is clearly characterised by a larger
5780/5797 \AA\ DIB ratio.

Another exciting thing revealed by this map is that within the Local Bubble
cavity exist discrete regions of a low 5780/5797 \AA\ DIB ratio; for instance
towards HD\,74521 at approx.\ ($-90$,$-70$) and ($-90$,$+70$) pc in the
Galactic Plane and meridional plane, or HD\,172910 at approx.\ ($+130$,$0$)
and ($+130$,$-40$) pc. These sight-lines must be hitting clumps of cold
neutral gas that somehow survive within the harsh environment of the Local
Bubble.

\section{Discussion}

\subsection{The carriers of the DIBs}

DIBs are currently thought to have molecular carriers, with the DIBs
representing vibronic transitions from a large number of different molecules
(McCall et al.\ 2010). All molecules in a given sight-line that give rise to a
particular DIB are expected to be in their vibronic ground state, implying
that two or more DIBs arising from that molecule should be part of a vibronic
progression in the molecule; the relative strengths of the vibronic bands
should then be exactly the same from one sight-line to another (McCall et al.\
2010). Therefore, a tight correlation between two DIBs is a promising, albeit
insufficient, condition to identify DIBs with a common carrier (Josafatsson \&
Snow 1987). The problem is, of course, that these molecules may not be in the
ground vibration state. Differences in excitation of low-lying energy levels
might explain subtle differences between DIBs that on the whole behave
concurrently.

The observations presented here of tenuous, generally exposed gas within and
around the Local Bubble indicate that the 6196 and 6614 \AA\ DIBs may not form
as perfect a pair as recently suggested -- McCall et al.\ (2010) determined a
correlation coefficient of 0.99, while here we determine $r=0.89$ only. While
our measurements are compatible with those of McCall et al., we show clear
evidence of differing equivalent width ratios of these DIBs and we uncovered a
statistically significant break in the relation around
$EW(6614)\sim0.02$--0.04 \AA. Such behaviour at very low column densities may
hold clues to the precise relation between the carriers of these two DIBs.

The measurements presented here show correlations between the 5780 and 5797
\AA\ DIBs on the one hand, and the 6196 and 6614 \AA\ DIBs on the other. The
5780 \AA\ DIB correlates better with the 6196 \AA\ DIB than with the 6614 \AA\
DIB; the 5797 \AA\ DIB correlates better with the 6614 \AA\ DIB than the 5780
\AA\ DIB does. This suggests that the carrier of the 6614 \AA\ DIB is more
closely related to that of the 5797 \AA\ DIB than of the 5780 \AA\ DIB. The
5797 \AA\ DIB is also correlated with the 6196 \AA\ DIB, even better than the
5780 \AA\ DIB is. This suggests that the carrier of the 6196 \AA\ DIB has
similarities to those of the 5780 \AA\ DIB and -- especially -- 5797 \AA\ DIB.
But, as the 5780 and 5797 \AA\ DIB show different behaviour, the carrier of
the 6196 \AA\ DIB cannot be identical to both the carrier of the 5797 \AA\ DIB
and that of the 5780 \AA\ DIB. In figure 12a the 6614/6196 \AA\ DIB ratio is
plotted against the 5780/5797 \AA\ DIB ratio. While the 6614/6196 \AA\ DIB
ratio does vary between different sight-lines, overall there is no
correspondence to the behaviour of the 5780/5797 \AA\ DIB ratio. The latter is
clearly high for low Na$^0$ column densities but the 6614/6196 \AA\ DIB ratio
is no different from that in denser columns. As in figure 6c, there is also a
clear lower limit to the  5780/5797 \AA\ DIB ratio $\sim2$, and a similar
lower limit is seen for the 6614/6196 \AA\ DIB ratio.

Intriguingly, the 6203/6196 \AA\ DIB ratio {\it does} correlate with the
5780/5797 \AA\ DIB ratio (Fig.\ 12b). This is especially convincing for the
denser columns, with $EW($Na\,{\sc i} D$_2)>0.2$ \AA. The 6203/6196 \AA\ DIB
ratio seems to drop in more tenuous columns, with $EW($Na\,{\sc i}
D$_2)<0.2$ \AA\ (around $EW(5780)/EW(5797)\approx 5$); perhaps this may be
understood if the 6196 \AA\ DIB is more readily formed than the 5797 \AA\ DIB
is (cf.\ Figs.\ 3e and 6) -- as long as the radiation field is not inhibitive.

We confirm that the 5780 \AA\ DIB favours weakly ionized environments and the
5797 \AA\ DIB favours neutral environments, with the relation between the two
breaking down above $EW(5797)>0.04$ \AA\ at high statistical significance:
those denser clouds are dominated by neutral matter and the skin no longer
grows in proportion. Compared to Na\,{\sc i}, the 5780 and 5797 \AA\ DIBs are
weaker at higher Galactic latitude. This suggests that the carriers are eroded
under the particularly harsh extra-planar conditions. Also, at $|b|>40^\circ$
we could not detect both 6614 and 6196 \AA\ DIBs.

%
\begin{figure}
\centerline{\vbox{
\epsfig{figure=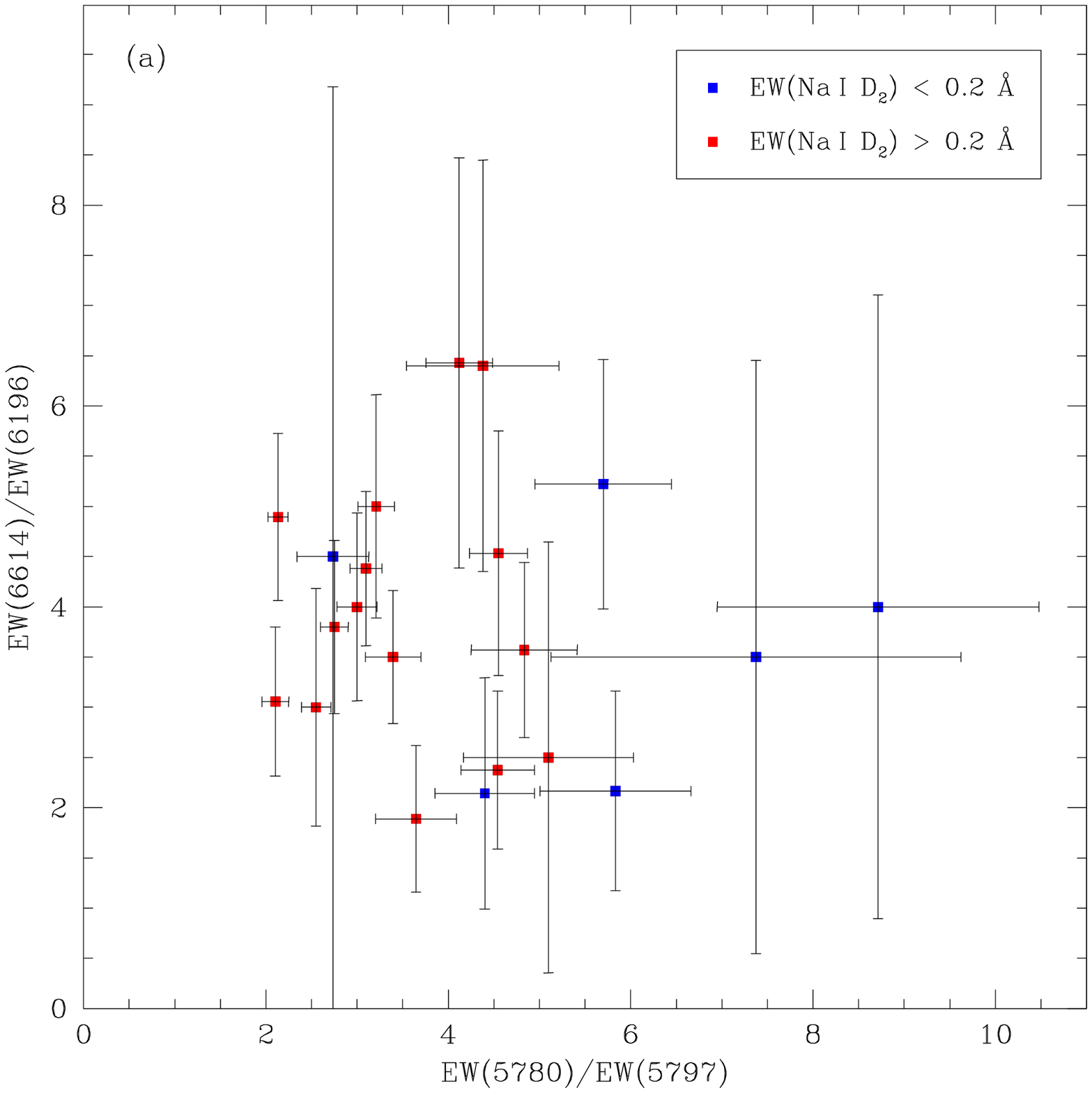,width=90mm}
\epsfig{figure=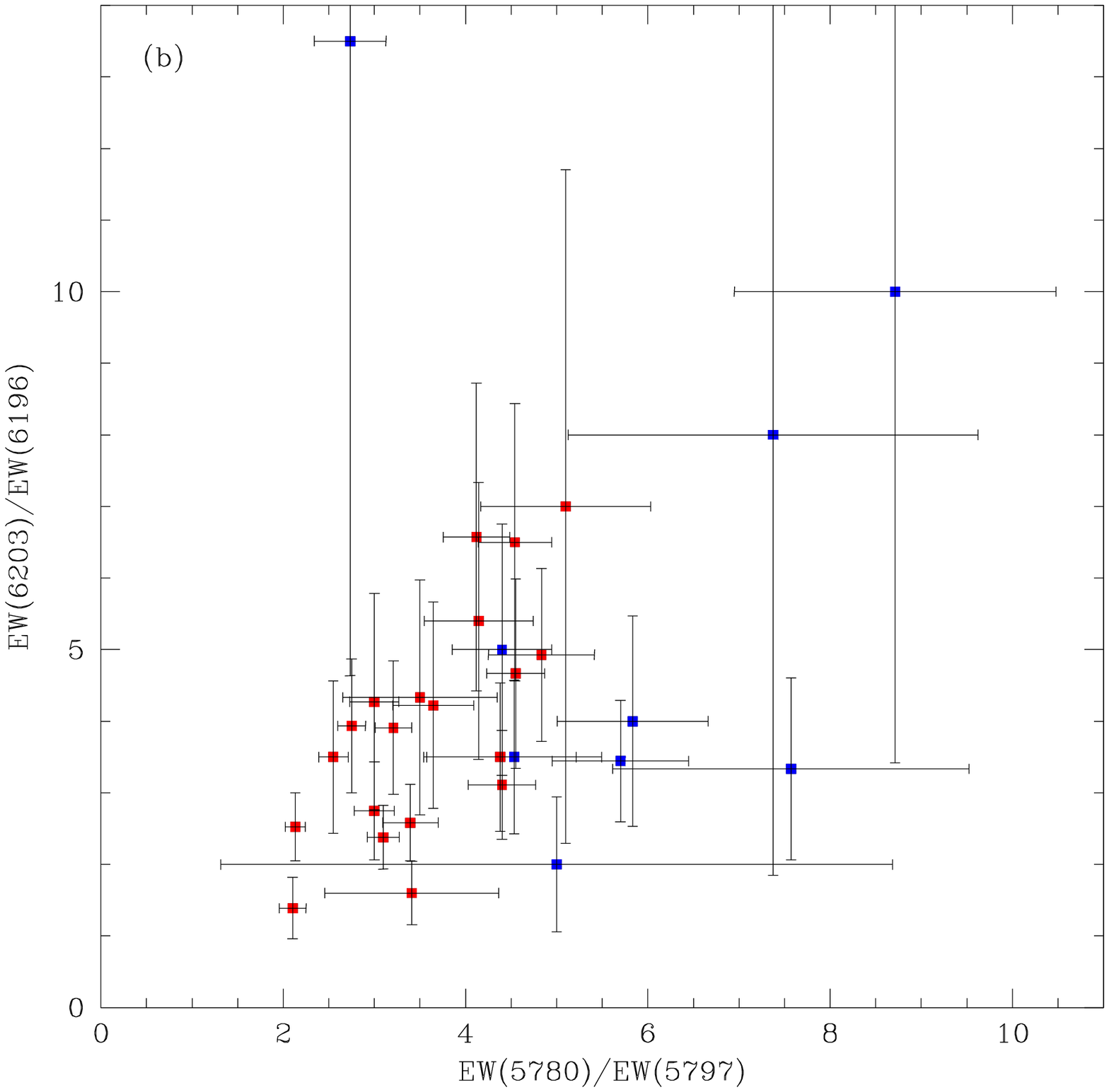,width=90mm}
}}
\caption[]{The 6614 to 6196 \AA\ DIB ratio vs.\ 5780 to 5797 \AA\ DIB ratio.
Distinction is made on the basis of the Na\,{\sc i} D$_2$ absorption; these
four DIBs were never {\it all} detected at high Galactic latitudes
($|b|>40^\circ$).}
\end{figure}

\subsection{Implications for our understanding of the Local Bubble}

The DIB maps of the Local Bubble presented in this work trace the cavity and
the wall of the Local Bubble rather well. The sight-lines with non-detections
are of equal importance as the sight-lines where DIBs are detected. In this
regard the Na\,{\sc i} map is less discriminative, as Na\,{\sc i} absorption
is nearly always detected and arises both in neutral and weakly-ionized gas.
The ``DIB-less'' sight-lines clearly reveal the extent of the cavity within
the Galactic Plane, as well as the vertical structure: the 5797 \AA\ DIB map
suggests a chimney opening out into the Halo though the 5780 \AA\ DIB is still
detected high up in the extra-planar gas. This difference is due to the
different alliance of these two DIBs to neutral and weakly-ionized gas,
respectively. The ``bursting'' of the Local Bubble has been noticed previously
(e.g., Lallement et al.\ 2003; Welsh et al.\ 1999, 2010; Welsh, Sallmen \&
Lallement 2004; Welsh \& Shelton 2009; Welsh \& Lallement 2012) but DIBs may
indicate an extra-planar interface between the Local Bubble and Halo.

Within the cavity, taking into account the distances to the targets, small
neutral cloudlets are found. So even if the Local Bubble is filled with hot
gas and/or the radiation field is unattenuated, denser, perhaps self-shielded
structures do exist within it. On the other hand, the ``wall'' of the Local
Bubble is porous, with some sight-lines piercing through it out to large
distances while adjacent sight-lines encounter much absorption. This must have
implications for the leakage of UV radiation -- and possibly hot gas -- out of
the Local Bubble into the surrounding ISM, both within the Halo and also the
Disc of the Milky Way.

\section{Summary}

In this first paper of a series we have presented the results from a Southern
hemisphere spectroscopic survey of 238 nearby early-type stars, aimed at
mapping the ISM within and surrounding the Local Bubble in the 5780 and 5797
\AA\ DIBs, and Na\,{\sc i}. We also investigated the relationship with other
DIBs detected in this spectral region, notably those at 6196 and 6614 \AA,
indicating subtle differences in their mutual behaviour. The DIB maps,
including detections as well as non-detections, clearly outline the extent of
the Local Bubble within the Galactic Plane, and its vertical structure opening
out into the Halo, and the generally harsh conditions within the Local Bubble
as indicated by a high 5780/5797 \AA\ DIB ratio throughout. However, small
more neutral cloudlets were seen close to the Sun as well, whilst the ``wall''
of the Local Bubble is far from impermeable. Thus we have demonstrated the
viability of using DIBs to map local ISM, and we will use these data in
combination with the complementary Northern hemisphere survey to reconstruct a
full three-dimensional map of the Local Bubble.

\begin{acknowledgements}
We wish to thank all of the staff at La Silla, and the dog. We also thank the
referee for a positive report which has helped improve the presentation of our
results. MB acknowledges an STFC studentship at Keele University. PJS thanks
the Leverhulme Trust for award of a Research Fellowship and the Leiden
Observatory for hosting an extended visit. This research has made use of the
SIMBAD database, operated at CDS, Strasbourg, France.
\end{acknowledgements}


\begin{thebibliography}{}
\bibitem[Bailey (2014)]{Bailey2014} Bailey A., 2014, Ph.D.\ thesis, Keele
University
\bibitem[Bailey et al.(2014)]{Bailey2014} Bailey M., van Loon J.Th., Sarre
P.J., Beckman J.E., 2015, arXiv:1509.05319
\bibitem[Baron et al.(2015)]{Baron2015} Baron D., Poznanski D., Watson D., Yao
Y., Prochaska J.X., 2015, MNRAS, 447, 454
\bibitem[Barstow et al.(2010)]{Barstow2010} Barstow M.A., Boyce D.D., Welsh
B.Y., Lallement R., Barstow J.K., Forbes A.E., Preval S., 2010, ApJ, 723, 1762
\bibitem[Breitschwerdt et al.(2009)]{Breitschwerdt2009} Breitschwerdt D., de
Avillez M.A., Fuchs B., Dettbarn C., 2009, Space Science Reviews, 143, 263
\bibitem[Cami et al.\ (1997)]{Cami1997} Cami J., Sonnentrucker P., Ehrenfreund
P., Foing B.H., 1997, A\&A, 326, 822
\bibitem[Cordiner et al.(2013)]{Cordiner2006} Cordiner M.A., Fossey S.J.,
Smith A.M., Sarre P.J., 2013, ApJ, 764, L10
\bibitem[Cox \& Anderson (1982)]{Cox1982} Cox D.P., Anderson P.R., 1982, ApJ,
253, 268
\bibitem[Cox (2005)]{Cox2005} Cox D.P., 2005, ARA\&A, 43, 337
\bibitem[Cox et al.(2006)]{Cox2006} Cox N.L.J., Cordiner M.A., Cami J., Foing
B.H., Sarre P.J., Kaper L., Ehrenfreund P., 2006, A\&A, 447, 991
\bibitem[Farhang et al.(2015a)]{Farhang2015a} Farhang A., Khosroshahi H.G.,
Javadi A., et al., 2015a, ApJ, 800, 64 (Paper II)
\bibitem[Farhang et al.(2015b)]{Farhang2015b} Farhang A., Khosroshahi H.G.,
Javadi A., van Loon J.Th., 2015b, ApJS, 216, 33 (Paper III)
\bibitem[Fitzgerald (1968)]{Fitzgerald1968} Fitzgerald M.P., 1968, AJ, 73, 983
\bibitem[Friedman et al.(2011)]{Friedman2011} Friedman S.D., York D.G., McCall
B.J., et al., 2011, ApJ, 727, 33
\bibitem[Frisch (1981)]{Frisch1981} Frisch P.C., 1981, Nature 293, 377
\bibitem[Frisch (1998)]{Frisch1998} Frisch P.C., 1998, in: ``The Local Bubble
and Beyond'', IAU Colloquium 166, eds.\ D.\ Breitschwerdt, M.J.\ Freyberg \&
J.\ Tr\"umper, Lecture Notes in Physics (Springer-Verlag), 506, p269
\bibitem[Galeazzi et al.(2014)]{Galeazzi2014} Galeazzi M., Chiao M., Collier
M.R., et al., 2014, Nature, 512, 171
\bibitem[Heger (1922)]{Heger1922} Heger M.L., 1922, Lick Observatory Bulletin,
10, 141
\bibitem[Herbig (1975)]{Herbig1975} Herbig G.H., 1975, ApJ, 196, 129
\bibitem[Herbig (1995)]{Herbig1995} Herbig G.H., 1995, ARA\&A, 33, 19
\bibitem[Hobbs et al.(2008)]{Hobbs2008} Hobbs L.M., York D.G., Snow T.P., et
al., 2008, ApJ, 680, 1256
\bibitem[Horne (1986)]{Horne1986} Horne K., 1986, PASP, 98, 609
\bibitem[Hurwitz, Sasseen \& Sirk (2005)]{Hurwitz2005} Hurwitz M., Sasseen
T.P., Sirk M.M., 2005, ApJ, 623, 911
\bibitem[Jenniskens \& D\'esert (1994)]{Jenniskens1994} Jenniskens P.,
D\'esert F.-X., 1994, A\&AS, 106, 39
\bibitem[Josafatsson \& Snow (1987)]{Josafatsson1987} Josafatsson K., Snow
T.P., 1987, ApJ, 319, 436
\bibitem[Kos et al.(2014)]{Kos2014} Kos J., Zwitter T., Wyse R., et al., 2014,
Science, 345, 791
\bibitem[Kre{\l}owski \& Westerlund (1988)]{Krelowski1988} Kre{\l}owski J.,
Westerlund B.E., 1988, A\&A, 190, 339
\bibitem[Lallement et al.(2003)]{Lallement2003} Lallement R., Welsh B.Y.,
Vergely J.L., Crifo F., Sfeir D., 2003, A\&A, 411, 447
\bibitem[Lallement et al.(2014a)]{Lallement2014a} Lallement R., Vergely J.-L.,
Valette B., Puspitarini L., Eyer L., Casagrande L., 2014a, A\&A, 561, 91
\bibitem[Lallement et al.(2014b)]{Lallement2014b} Lallement R., Bertaux J.-L.,
Qu\'emerais E., Sandel B.R., 2014b, A\&A, 563, 108
\bibitem[Lucke (1978)]{Lucke1978} Lucke P.B., 1978, A\&A, 64, 367
\bibitem[Ma\'{\i}z-Apell\'aniz (2001)]{MaizApellaniz2001}
Ma\'{\i}z-Apell\'aniz J., 2001, ApJ, 560, L83
\bibitem[McCall et al.(2010)]{McCall2010} McCall B.J., Drosback M.M., Thorburn
J.A., et al., 2010, ApJ, 708, 1628
\bibitem[Merrill \& Wilson (1938)]{Merrill1938} Merrill P.W., Wilson O.C.,
1938, ApJ, 87, 9
\bibitem[Meyer et al.(2012)]{Meyer2012} Meyer D.M., Lauroesch J.T., Peek
J.E.G., Heiles C., 2012, ApJ, 752, 119
\bibitem[Peek et al.(2011)]{Peek2011} Peek J.E.G., Heiles C., Peek K.M.G.,
Meyer D.M., Lauroesch J.T., 2011, ApJ, 735, 129
\bibitem[Puspitarini, Lallement \& Chen (2013)]{Puspitarini2013} Puspitarini
L., Lallement R., Chen H.-C., 2013, A\&A, 555, 25
\bibitem[Raimond et al.(2012)]{Raimond2012} Raimond S., Lallement R., Vergely
J.L., Babusiaux C., Eyer L., 2012, A\&A, 544, 136
\bibitem[Sarre (2006)]{Sarre2006} Sarre P.J., 2006, Journal of Molecular
Spectroscopy, 238, 1
\bibitem[Smith et al.(2013)]{Smith2013} Smith K.T., Fossey S.J., Cordiner
M.A., Sarre P.J., Smith A.M., Bell T.A., Viti S., 2013, MNRAS, 429, 939
\bibitem[Snowden et al.(1998)]{Snowden1998} Snowden S.L., Egger R., Finkbeiner
D.P., Freyberg M.J., Plucinsky P.P., 1998, ApJ, 493, 715
\bibitem[Snowden et al.(2015)]{Snowden2015} Snowden S.L., Koutroumpa D., Kuntz
K.D., Lallement R., Puspitarini L., 2015, ApJ, 806, 120
\bibitem[Tielens (2014)]{Tielens2014} Tielens A.G.G.M., 2014, in: ``The
Diffuse Interstellar Bands'', IAU Symposium 297, eds.\ J.\ Cami \& N.L.J.\ Cox
(Cambridge University Press), p399
\bibitem[van Leeuwen (2007)]{vanLeeuwen2007} van Leeuwen F., 2007, A\&A, 474,
653
\bibitem[van Loon et al.(2009)]{vanLoon2009} van Loon J.Th., Smith K.T.,
McDonald I., Sarre P.J., Fossey S.J., Sharp R.G., 2009, MNRAS, 399, 195
\bibitem[van Loon et al.(2013)]{vanLoon2013} van Loon J.Th., Bailey M., Tatton
B.L., et al., 2013, A\&A, 550, 108
\bibitem[Welsh et al.(1999)]{Welsh1999} Welsh B.Y., Sfeir D.M., Sirk M.M.,
Lallement R., 1999, A\&A, 352, 308
\bibitem[Welsh, Sallmen \& Lallement (2004)]{Welsh2004} Welsh B.Y., Sallmen
S., Lallement R., 2004, A\&A, 414, 261
\bibitem[Welsh \& Shelton (2009)]{Welsh2009} Welsh B.Y., Shelton R.L., 2009,
Astrophysics and Space Science, 323, 1
\bibitem[Welsh et al.(2010)]{Welsh2010} Welsh B.Y., Lallement R., Vergely
J.-L., Raimond S., 2010, A\&A, 510, 54
\bibitem[Welsh \& Lallement (2012)]{Welsh2012} Welsh B.Y., Lallement R., 2012,
PASP, 124, 566
\bibitem[Welty \& Hobbs (2001)]{Welty2001} Welty D.E., Hobbs L.M., 2001, ApJS,
133, 345
\bibitem[Wenger et al.(2000)]{Wenger2000} Wenger M., Ochsenbein F., Egret D.,
et al., 2000, A\&AS, 143, 9
\bibitem[Westerlund \& Kre{\l}owski (1988)]{Westerlund1998} Westerlund B.E.,
Kre{\l}owski J., 1988, A\&A, 203, 134
\bibitem[Zasowski et al.(2015)]{Zasowski2015} Zasowski G., M\'enard B.,
Bizyaev D., et al., 2015, ApJ, 798, 35
\bibitem[Zirnstein, Heerikhuisen \& McComas (2015)]{Zirnstein2015} Zirnstein
E.J., Heerikhuisen J., McComas D.J., 2015, ApJ, 804, L22
\end{thebibliography}
\end{document}